\documentclass[fleqn,usenatbib,linenumbers]{mnras}
\usepackage{eso-pic}
\AddToShipoutPictureBG*{%
  \AtPageUpperLeft{%
    \hspace{0.75\paperwidth}%
    \raisebox{-3.5\baselineskip}{%
      \makebox[0pt][l]{\textnormal{DES-2015-0048}}
}}}%

\AddToShipoutPictureBG*{%
  \AtPageUpperLeft{%
    \hspace{0.75\paperwidth}%
    \raisebox{-4.5\baselineskip}{%
      \makebox[0pt][l]{\textnormal{FERMILAB-PUB-20-655-AE}}
}}}%

\usepackage{lineno}
\usepackage{newtxtext,newtxmath}
 
\usepackage{mdframed}
\usepackage{graphicx}
\usepackage{subcaption}
	
\usepackage{lineno}

\usepackage{hyperref}
\usepackage[T1]{fontenc}
\usepackage{ae,aecompl}
\usepackage{todonotes}

\usepackage{graphicx}	
\usepackage{amsmath}	
\usepackage{amssymb}	
\usepackage{adjustbox}
\newcommand{\Nz}{$n(z)$}
\newcommand{\Nuz}{$n_{{\rm u}}(z)$}
\newcommand{\Nrz}{$n_{{\rm r}}(z)$}
\newcommand{\phiu}{{n_{\rm u}}}
\newcommand{\phir}{{n_{\rm r}}}

\newcommand{\bu}{b_{\rm{u}}}

\newcommand{\wur}{\bar {w}_{\rm{ur}}}

\newcommand{\redmagic}{\textit{redMaGiC}}

\newcommand{\avg}[1]{\langle #1 \rangle}

\newcommand{\mcal}{\textsc{metacalibration}}

\newcommand{\vecp}{\ensuremath{\mathbf{p}}}
\newcommand{\vecq}{\ensuremath{\mathbf{q}}}
\newcommand{\vecs}{\ensuremath{\mathbf{s}}}
\DeclareMathOperator{\sys}{Sys}

\newcommand{\mcalR}{\mbox{\boldmath $R$}}

\newcommand{\mcalRg}{\mbox{\boldmath $R_\gamma$}}
\newcommand{\mcalRs}{\mbox{\boldmath $R_s$}}

\newcommand\gbso[1]{}
\newcommand{\gary}[1]{\textcolor{black}{#1}}
\newcommand{\giulia}[1]{\textcolor{black}{#1}}
\newcommand{\marco}[1]{\textcolor{black}{#1}}

\title[Clustering Redshifts]{Dark Energy Survey Year 3 Results: Clustering Redshifts -- Calibration of the Weak Lensing Source Redshift Distributions with \redmagic\ and  BOSS/eBOSS}

\author[Gatti \& Giannini et~al.]{
\parbox{\textwidth}{
\Large 
M. Gatti$^{1,\star}$, 
G. Giannini$^{1,\dag}$, 
G.~M.~Bernstein$^{2}$, 
A.~Alarcon$^{3}$, 
J. Myles$^{4}$, 
A.~Amon$^{5}$,
R.~Cawthon$^{6}$,
M. Troxel$^{7}$, 
J. DeRose$^{8,9}$, 
S.~Everett$^{9}$,
A.~J.~Ross$^{10}$,
E.~S.~Rykoff$^{5,11}$,
J.~Elvin-Poole$^{10,12}$,
J.~Cordero$^{13}$,
I. Harrison$^{13}$, 
C. Sanchez$^{2}$, 
J.~Prat$^{14}$,
D.~Gruen$^{4,5,11}$,
H. Lin$^{15}$, 
M. Crocce$^{16,17}$, 
E. Rozo$^{18}$,
T.~M.~C.~Abbott$^{19}$,
M.~Aguena$^{20,21}$,
S.~Allam$^{15}$,
J.~Annis$^{15}$,
S.~Avila$^{22}$,
D.~Bacon$^{23}$,
E.~Bertin$^{24,25}$,
D.~Brooks$^{26}$,
D.~L.~Burke$^{5,11}$,
A.~Carnero~Rosell$^{27,28}$, 
M.~Carrasco~Kind$^{29,30}$, 
J.~Carretero$^{1}$,
F.~J.~Castander$^{16,17}$,
A.~Choi$^{10}$,
C.~Conselice$^{13,31}$,
M.~Costanzi$^{32,33}$,
M.~Crocce$^{16,17}$,
L.~N.~da Costa$^{21,34}$, 
M.~E.~S.~Pereira$^{35}$,
K.~Dawson$^{36}$,
S.~Desai$^{37}$,
H.~T.~Diehl$^{15}$,
K.~Eckert$^{2}$,
T.~F.~Eifler$^{38,39}$,
A.~E.~Evrard$^{40,35}$,
I.~Ferrero$^{41}$,
B.~Flaugher$^{15}$,
P.~Fosalba$^{16,17}$, 
J.~Frieman$^{15,42}$, 
J.~Garc\'ia-Bellido$^{22}$, 
E.~Gaztanaga$^{16,17}$, 
T.~Giannantonio$^{43,44}$, 
R.~A.~Gruendl$^{29,30}$,
J.~Gschwend$^{21,34}$ 
S.~R.~Hinton$^{45}$,
D.~L.~Hollowood$^{9}$,
K.~Honscheid$^{10,12}$,
B.~Hoyle$^{46,47,48}$,
D.~Huterer$^{34}$,
D.~J.~James$^{49}$,
K.~Kuehn$^{50,51}$,
N.~Kuropatkin$^{15}$,
O.~Lahav$^{26}$,
M.~Lima$^{20,21}$,
N.~MacCrann$^{52}$,
M.~A.~G.~Maia$^{21,34}$,
M.~March$^{2}$,
J.~L.~Marshall$^{53}$,
P.~Melchior$^{54}$,
F.~Menanteau$^{29,30}$,
R.~Miquel$^{55,1}$,
J.~J.~Mohr$^{46,47}$,
R.~Morgan$^{6}$,
R.~L.~C.~Ogando$^{21,33}$,
A.~Palmese$^{15,42}$,
F.~Paz-Chinch\'{o}n$^{43,30}$,
W. J. Percival$^{56,57,58}$,
A.~A.~Plazas$^{53}$,
M.~Rodriguez-Monroy$^{59}$,
A.~Roodman$^{5,11}$,
G. Rossi$^{60}$,
S.~Samuroff$^{61}$,
E.~Sanchez$^{59}$,
V.~Scarpine$^{15}$,
L.~F.~Secco$^{2}$,
S.~Serrano$^{16,17}$,
I.~Sevilla-Noarbe$^{59}$,
M.~Smith$^{62}$,
M.~Soares-Santos$^{34}$,
E.~Suchyta$^{63}$,
M.~E.~C.~Swanson$^{30}$,
G.~Tarle$^{35}$,
D.~Thomas$^{23}$,
C.~To$^{4,5,11}$,
T.~N.~Varga$^{47,48}$,
J.~Weller$^{47,48}$,
R.D.~Wilkinson$^{64}$ 
\begin{center} (DES Collaboration) \end{center}
}
}

\vspace{0.2cm}

\date{Accepted XXX. Received YYY; in original form ZZZ}

\pubyear{2018}

\begin{document}
\label{firstpage}
\pagerange{\pageref{firstpage}--\pageref{lastpage}}
\maketitle

\begin{abstract}
We present the calibration of the Dark Energy Survey Year 3 (DES Y3) weak lensing source galaxy redshift distributions {$n(z)$} from clustering measurements. In particular, we cross-correlate the weak lensing (WL) source galaxies sample with \redmagic\  galaxies (luminous red galaxies with secure photometric redshifts) and a spectroscopic sample from BOSS/eBOSS to estimate the redshift distribution of the DES sources sample. \gbso{The recovered redshift distributions from clustering are used as a prior for the fiducial photo-$z$ redshift distribution posteriors used in the main cosmological analysis. The method presented here includes several upgrades compared to the DES Y1 procedure, including a treatment of the galaxy-matter bias evolution.} {Two distinct methods for using the clustering statistics are described.  The first uses the clustering information independently to estimate the mean redshift of the source galaxies within a redshift window, as done in the DES Y1 analysis.  The second method establishes a likelihood of the clustering data as a function of $n(z)$, which can be incorporated into schemes for generating samples of $n(z)$ subject to combined clustering and photometric constraints.  Both methods incorporate marginalization over various astrophysical systematics, including magnification and redshift-dependent galaxy-matter bias.}  We characterize the \gbso{systematic} uncertainties of the methods in simulations; the first method recovers the mean $z$ of tomographic bins to RMS (precision) of $\sim 0.014$.  Use of the second method is shown to vastly improve the accuracy of the shape of $n(z)$ derived from photometric data. \gbso{showing the accuracy of the method  to be X, Y\% better than DES Y1} The two methods are then applied to the DES Y3 data.
\end{abstract}

\begin{keywords}
 galaxies: distances and redshifts -- cosmology: observations 
\end{keywords}

\makeatletter
\def \blfootnote{\xdef\@thefnmark{}\@footnotetext}
\makeatother

\setcounter{footnote}{1}
\blfootnote{$^{\star}$ E-mail: mgatti@ifae.es}
\blfootnote{$^{\dag}$ E-mail: ggiannini@ifae.es}



\section{Introduction}

The Dark Energy Survey (DES) is a photometric survey that has imaged 5,000 deg$^2$ of the sky. The DES Y3 '3x2' analysis \citep{y3-3x2ptkp} using data taken during the first 3 seasons of observations constrains cosmological parameters by combining three different measurements of two-point correlation functions: cosmic shear \citep{y3-cosmicshear1,y3-cosmicshear2}, galaxy-galaxy lensing \citep{y3-gglensing}, and galaxy clustering \citep{y3-galaxyclustering}. The cosmic shear measurement probes the angular correlation of more than 100,000,000 galaxy shapes from the weak lensing sample \citep*{y3-shapecatalog}, divided into four tomographic bins. The cross-correlation of galaxy shapes and the positions of red luminous galaxies identified by the \redmagic\ algorithm \citep{Rozo2016} is measured by galaxy-galaxy lensing. Lastly, galaxy clustering measures the auto-correlation of the positions of \redmagic\ galaxies. A magnitude-limited sample \citep{y3-2x2maglimforecast} will be also used as lens sample alternatively to \redmagic\ in a second analysis \citep{y3-2x2ptaltlensresults}, with the goal of improving the cosmological constraints.


The correct cosmological interpretation of these measurements relies on an accurate estimate of the redshift distributions of the samples, which can otherwise lead to biases in the inferred cosmological parameters \citep[e.g.][]{Huterer2006,Hildebrandt2012,Choi2016,Hoyle2018}. Photometric surveys have been relying on different methodologies to derive redshift distributions \citep{Hildebrandt2010,Sanchez2014}, mostly based on galaxies' multi-band photometry (photo-$z$ methods, {or PZ}). However, these methods are
ultimately limited by {the redshift ambiguities in few-band colors, and the limited and incomplete spectroscopic samples available to calibrate the color-redshift relations.} \gbso{the fact that any mapping from a set of photometric observables to redshift can be ambiguous.} 

Clustering-based redshift methods (\citealt{Newman2008,Menard2013,Davis2017aaa,Morrison2017,Scottez2017,Johnson2017}; \citealt*{Gatti2018}; \citealt{vandenBusch2020,Hildebrandt2020}) \gbso{met with success in the past years providing alternative ways with respect} {offer an alternative} to standard photo-$z$ methods to infer redshift distributions. In short, clustering-based methods exploit the two-point correlation signal between a photometric ``unknown'' sample and a ``reference'' sample of high-fidelity redshift galaxies divided into thin bins, to infer the redshift distributions of the photometric sample. One of the biggest advantages of clustering-based methods is that the reference sample does not have to be representative of the photometric sample. Clustering-based methods (or clustering-$z$, or WZ) have been in the past years successfully applied to both data \citep{Hildebrandt2017,Johnson2017,Davis2017aaa,Davis2018,Cawthon2018,Bates2019,vandenBusch2020,Hildebrandt2020} and simulations (\citealt{Schmidt2013,McQuinn2013,Scottez2017}; \citealt*{Gatti2018}), and they represent one credible \gbso{alternative} {supplement} to standard photo-$z$ methods for the new, upcoming generation of data sets \citep{Scottez2017}. 

\gbso{Depending on the particular application, cross-correlation methods were used} {Cross-correlation methods have been used both} to provide an independent redshift distribution estimate \gbso{or they have been used} and to calibrate \gbso{other} distributions inferred from photo-$z$ methods. In the DES Y1 cosmological analysis we opted for the latter approach (\citealt{Davis2017aaa}; \citealt*{Hoyle2018}). In particular, we used high quality photometric redshifts provided by \redmagic\ galaxies \citep{Rozo2016} to measure the clustering-$z$ signal with the weak lensing (WL) source-galaxy sample. The \gbso{choice of using}{use of} high quality photometric redshifts rather than spectroscopic redshifts was motivated by the higher statistical power of the \redmagic\ sample, owing to the large number of \redmagic\ galaxies ($650,000$ for DES Y1) in the DES footprint. Due to the limited redshift range of the \redmagic\ sample, clustering-$z$ estimates \gbso{could not have been used on their own,}{could not have been used to determine $n(z)$ in its entirety on their own,} but they \gbso{have been used to}{have been used to} calibrate the mean redshift of the distributions measured by other DES photo-$z$ methods {(with the mean taken over the \redmagic\ $z$ bounds).} A similar approach has been implemented by the KiDS team in their recent cosmological analysis \citep{vandenBusch2020,Hildebrandt2020}, where they used cross-correlation estimates to calibrate the mean redshifts inferred from other photo-$z$ methods. They used a number of different spectroscopic samples as a reference sample, which guaranteed a greater redshift coverage but less statistical power compared to the use of \redmagic\ galaxies.

The strategy for calibration of the WL redshift distributions for DES Y3 improves in multiple respects on the Y1 strategy outlined in \cite*{Gatti2018}. From the clustering-redshift side, we \gbso{explore} {execute} two different methods to {combine clustering information with} \gbso{calibrate} redshift distributions from {photometry.} \gbso{other photo-$z$ codes using the clustering information: a method that mostly focused on calibrating} {The first approach is to use clustering-$z$ to estimate the mean redshift $\langle z \rangle_{\rm wz}$, and assign a clustering-$z$ likelihood to any candidate $n(z)$ from photo-$z$ techniques based on the value of its mean $\langle z \rangle_{\rm pz}$} \gbso{the mean redshift of the distributions} (similar to the DES Y1 analysis).  {We will refer to this as the ``mean-matching'' approach. The second, new method is to pose both the clustering-$z$ and the photo-$z$ measurements as probabilities $p[D | n(z)]$ of the observational data $D$ given redshift distributions $n(z)$; then to sample the full $n(z)$ from the posterior $p[n(z)]$ implied by multiplying these probabilities. We will refer to this as the ``full-shape'' method.} \gbso{method that calibrates the full shape of the distributions.}

We furthermore improve {over Y1 in} the modelling of the clustering signal, accounting for  the redshift evolution of the galaxy-matter bias and the clustering of the underlying dark matter density field, which were neglected in the DES Y1 analysis. In the second method that calibrates the shape of the redshift distributions, we also marginalise over magnification effects. Finally, we use a combination of two different reference samples: \redmagic\ galaxies with high-quality photometric redshifts; and a spectroscopic sample from the combined BOSS (Baryonic Oscillation Spectroscopic Survey, \citealt{boss}) and eBOSS (extended-Baryon Oscillation Spectroscopic Survey, \citealt{eboss2016,dr16,Alam2020}) catalogs.  Only \redmagic\ galaxies were used in DES Y1. On one hand, \redmagic\ galaxies span the full DES Y3 footprint \citep{y3-galaxyclustering} and are characterised by a higher number density than BOSS/eBOSS galaxies, which cover only $\approx17\%$ of the DES Y3 footprint. On the other hand, the latter sample spans a wider redshift range and has better redshift estimates, which makes the combination of the two samples desirable. 


The fiducial photo-$z$ estimates for the DES Y3 weak lensing sample are provided by a self-organizing-map-based scheme (hereafter SOMPZ,  \citealt{Buchs2019}; \citealt*{y3-sompz}). The SOMPZ method provides {a means to generate samples of the $n(z)$ for all tomographic bins that}
\gbso{for each tomographic bin a number of posterior samples of the redshift distribution which} encompass the uncertainties in the photometric inference of the distributions. \gbso{These realisations are sampled over when estimating the cosmological parameters in the fiducial DES Y3 cosmological analysis. In this context, we use clustering redshift estimates to provide priors for the SOMPZ realisations.}  {The mean-matching clustering-$z$ method may be used to confirm or adjust the $n(z)$ samples generated by SOMPZ.  We use the full-shape method as the fiducial method for DES Y3, generating samples of $n(z)$ from the combined SOMPZ and clustering likelihoods. In either route, the DES Y3 cosmological analysis  is done by sampling over the finite set of realizations generated by SOMPZ$+$clustering-z.}

We note that there exist other strategies to combine clustering-based and photo-$z$ estimates. For example, \cite{Sanchez2019} and \cite*{Alarcon2019} show how to combine photo-$z$ and clustering-based estimates using a hierarchical Bayesian model \citep{Leistedt2016}. The application of these methods to DES data is left for future work.

This paper is organised as follows. In Section~\ref{sect:methodology}  we describe the two different methodologies used in DES Y3 to calibrate photo-$z$ posteriors using clustering-based redshift estimation, and explain how to assign a likelihood to the cross-correlation information. The simulations and the data sets used in this paper are described and compared in Section~\ref{sect:data}. In Section~\ref{sect:sims_tests} we perform extended tests in simulations assessing the systematic uncertainty of the methods. The calibration on DES Y3 data is presented in Section~\ref{sect:calibration}, and in Section~\ref{sect:conclusions} we discuss future prospects for this method and present our conclusions.  

\section{Methodology}\label{sect:methodology}
{We describe the clustering-$z$ methodology as generally as possible in this section, deferring to \S\ref{sect:data} the description (and the choice of the binning) of the particular samples adopted for DES Y3.}
\gbso{In the DES Y3 analysis we will use clustering-based redshift estimates as a prior for the photo-$z$ posterior distributions of a given science sample. We defer the description (and the choice of the binning) of the particular samples adopted in this work to $\S$\ref{sect:data}, while keeping the description of the methodology as general as possible.}

%


\subsection{Modelling and measuring the correlation signal} \label{method_clustering-z}

Clustering-based methods rely on the assumption that the cross-correlation between two samples of objects is non-zero only in the case of overlap of the distribution of objects in physical space, due their mutual gravitational influence. Let us consider two samples:
\begin{enumerate}
  \item    An \textit{unknown} sample, whose redshift distribution \Nuz\  has to be measured, namely our WL source sample, and
  \item   A \textit{reference} sample, whose redshift distribution \Nrz\ is known (either from spectroscopic redshifts or from high-precision photometric redshifts). \gbso{The reference sample is divided into narrow redshift bins. }
 \end{enumerate}
\gbso{To calibrate the redshift distribution of the unknown sample we bin the reference sample in narrow redshift bins, and then}{We compute the angular clustering signal ${w}_{\rm ur}$ as a function of the separation angle $\theta$  between the unknown sample and the reference population.} \gbso{each of these reference redshift bins.} \marco{Under the assumption of linear biasing and the Limber approximation \citep{Limber}, the clustering signal can be written as (e.g., \citealt{methodpaper}):}
%
%
\begin{equation}
\label{crosscorr}
{w}_{{\rm ur}}(\theta) = \int  dz'\phiu(z')\phir(z')
 	b_{\rm u}(z')b_{\rm r}(z'){w}_{\rm DM}(\theta,z')  + M(\theta),
\end{equation}
where $\phiu(z')$ and $\phir(z')$ are the unknown- and reference-sample redshift distributions (normalised to unity over the full redshift interval), $b_{\rm u}(z')$ and  $b_{\rm r}(z')$  are the linear galaxy-matter biases of the two samples, and ${w}_{\rm DM}(\theta,z')$ is the dark-matter 2-point angular correlation function.  The term $M(\theta)$ refers to the contribution of lensing magnification effects; \giulia{description and full expressions for the terms} ${w}_{\rm DM}(\theta,z')$ and $M(\theta)$ {are detailed below (Eq. \ref{eq:clustzonly} and Eq. \ref{eq:magnification}).} \giulia{Note that while we acknowledge that the assumption of linear biasing is not expected to hold at small scales, we are nevertheless confident we are able to estimate the systematic bias introduced by this premise, as explained in Section \ref{sec:combination}}. We also note that the Limber approximation is a standard assumption in clustering redshift works, and it is expected to have a minimal impact on our results (e.g., \citealt{McQuinn2013}).


\gbso{In our methodology, we use a ``1-point estimate'' of the correlation function. In practice,} The correlation function is measured as a function of angle, and \gbso{the recovered correlation function is} averaged over angular scales {to produce a ``scalar'' value} via
\begin{equation}
\label{crosscor2}
\wur=\int_{\theta_{\min}}^{\theta_{\max}}d\theta\ W(\theta){w}_{\rm ur}(\theta) ,
\end{equation}
where $W(\theta) \propto \theta^{-\gamma}$ is a weighting function. 
We {adopt} $\gamma=1$ \gbso{to increase the S/N}{to yield optimal S/N on the scalar in the presence of shot noise.} The integration limits in the integral in Eq.~\ref{crosscor2}  correspond to fixed physical scales. We use the \cite{DavisPeebles1983} estimator for the cross-correlation signal,
\begin{equation}
\label{DavisPeeblesestimator}
w_{\rm ur} (\theta) =\frac{N_{\rm Rr}}{N_{\rm Dr}}\frac{D_{\rm u}D_{\rm r}(\theta)}{D_{\rm u}R_{\rm r}(\theta)}-1,
\end{equation}
where $D_{\rm u}D_{\rm r}(\theta)$ and $D_{\rm u}R_{\rm r}(\theta)$ are respectively data--data and data--random pairs.  The pairs are properly normalized through ${N_{\rm Dr}}$ and ${N_{\rm Rr}}$, corresponding to the total number of galaxies in the reference sample and in the reference random catalog. \marco{If weights for the reference catalog of galaxies (or for the catalog of randoms) are provided, ${N_{\rm Dr}}$ (or ${N_{\rm Rr}}$) is the sum of the weights of the catalog, and $D_{\rm u}D_{\rm r}(\theta)$ (or $D_{\rm u}R_{\rm r}(\theta)$) is the weighted number of pairs. Note that weights can also be assigned to the unknown sample; in that case, the weighted number of pairs $D_{\rm u}D_{\rm r}(\theta)$ (or $D_{\rm u}R_{\rm r}(\theta)$) also accounts for the weights of the unknown sample.} As in \cite*{Gatti2018}, we use the  Davis \& Peebles estimator rather than the \cite{LandySzalay1993} estimator since the former involves using a catalog of random points for just one of the two samples. This allows us to avoid creating high-fidelity random catalogs for the DES Y3 source galaxy sample, whose selection function is very complex and non-trivial to replicate, besides being computationally very costly. For our analysis, we only rely on random points for the reference sample, whose selection function and mask are well understood.
\marco{We note that in the rest of the paper we adopted the Davis \& Peebles estimator even when measuring the auto-correlation of the reference samples, but we checked that using the {Landy \& Szalay} estimator lead to negligible variations.}

{Now we assume that}\gbso{Assuming the} the reference sample is divided into redshift bins centered at $z_i$, {each narrow enough that we can approximate $ \phir_{,i}(z)  \approx \delta_D (z-z_i)$ (with $\delta_D $ being Dirac's delta distribution and the integrands in Eq.~\ref{crosscorr} other than $n_{\rm r}$ can be treated as constant.} \gbso{. Hence,} Eqs.~\ref{crosscorr} and \ref{crosscor2} become:
\begin{equation}
\label{menard2}
\wur(z_i) \approx  n_{\rm u}(z_i) b_{\rm u}(z_i) b_{\rm r}(z_i)
 \bar{w}_{\rm DM}(z_i) + \bar{M}(z_i),
\end{equation}
where barred quantities indicate they have been averaged over angular scales {as per Eq.~\ref{crosscor2}.} \gbso{, reflecting the fact that we are using 1-angular bin estimates of the correlation while weighting pairs by their inverse separation.} In what follows we will, for simplicity, drop the bar. The above quantity is always estimated at the redshift $z_i$ of the $i$-th thin reference sample bin.

{The goal is to use Eq.~\ref{menard2} to infer $n_{\rm u}(z),$ the unknown redshift distribution, from the multiple measures $w_{\rm ur}(z_i).$  But it is important to note that this equation follows from a simplifying assumption.}
\gbso{ We also note that} We assumed the galaxy-matter bias to be described by a single number at all scales; this is true at large scales in the linear regime, but we do not expect this to necessary hold at the small scales used in this work. In the non-linear regime, even the fact that the terms inside the integral factorizes into ${b}_{\rm r}(z_i)  {b}_{\rm u}(z_i){w}_{\rm DM}(z_i)$ is not guaranteed \citep{Bernardeau2002,Desjacques2018}. \gbso{and can introduce a small systematic uncertainty.} The linear-bias assumption introduces a small systematic uncertainty that needs to be quantified.


{The evolution of} the quantities $b_{\rm r}(z_i)$,  $b_{\rm u}(z_i)$, ${w}_{\rm DM}(z_i)$,
${M}(z_i)$ \gbso{evolve with redshift, they} need to be characterized to correctly recover the redshift distribution of the unknown sample. We turn now to how to model or estimate these terms.

\begin{itemize}
    \item \textbf{The galaxy-matter bias evolution of the reference sample $b_{\rm r}(z)$}. As long as the redshifts of the reference sample are accurate enough, {and we assume linear biasing,} we can estimate $b_{\rm r}(z)$ by measuring the \gbso{1-point} {angle-averaged} estimate of the auto-correlation function of the reference sample divided into thin redshift bins ($\delta z = 0.02$) centered at $z_i$:
    \begin{equation}
    \label{autocorrref22}
        {w}_{\rm rr}(z_i) = \int dz'   \left[b_{\rm r}(z') n_{\rm r,i}(z')\right]^2 {w}_{\rm DM}(z').
    \end{equation}
    If the bins are sufficiently narrow so as to consider the biases and ${w}_{\rm DM}$ constant over the distributions, they can be pulled out of the above integrals.
    \begin{equation}
    \label{autocorrelation1}
        {w}_{\rm rr}(z_i) =  b_{\rm r}^2 (z_i) {w}_{\rm DM}(z_i) \int dz'  n_{\rm r,i}^2(z'), 
    \end{equation}    
    Knowledge of the redshift distributions of the narrow bins is then required to use Eq.~\ref{autocorrelation1} to estimate $b_{\rm r} (z_i)$. Lastly, we need to model ${w}_{\rm DM}(z)$ to correctly recover $b_{\rm r} (z)$.
    
  \item \textbf{The galaxy-matter bias evolution of the unknown sample $b_{\rm u}(z)$}. In principle, the auto-correlation of the unknown sample constrains this.  However in our case, $\phiu(z)$ is broad and unknown, and $b_{\rm u}$ likely varies substantially across the sample, so the information on $b_{\rm u}$ from the auto-correlation is weak and entangled with $\phiu$ itself.  The degeneracy between $b_{\rm u}$ and $n_{\rm u}$ is the fundamental limiting factor of clustering-$z$ methods. \giulia{While} mitigation schemes exist (e.g., \citealt{MattewsNewman2010,vandenBusch2020}), they are not free from shortcomings, so we decided not to attempt correcting for $b_{\rm u}$.
\gbso{it could be estimated similarly to the bias of the reference sample. The limiting factor that prevents us from using the auto-correlation functions to estimate the galaxy--matter biases evolution for the unknown sample is the poor photo-$z$ quality of the unknown sample.} \marco{Since it is difficult to place a priori constraints on $b_{\rm u}$, when forward modelling the clustering signal we chose to parameterize it in a flexible way (see \S \ref{sect:fsmethod}), effectively treating it as a free function.}

    \item \textbf{The dark matter 2-point correlation function  ${w}_{\rm DM}(z)$}. This can be modeled assuming a given cosmology and a non-linear power spectrum. At fixed $z_i$, this can be written as:
    

\begin{multline}
\label{eq:clustzonly}
{w}_{\rm DM}(z_i)=  \int d\theta W(\theta) \sum \frac{2 \ell +1}{4 \pi}P_{\ell}({\rm\cos} \theta)\\
\frac{1}{\chi(z_i)^2H(z_i)} P_{\rm NL}\left( \frac{l+1/2}{\chi(z_i)},z_i\right),
\end{multline}
 where $\chi$ is the comoving distance, $H(z_i)$ is the Hubble expansion rate at redshift $z_i$. $P_{\ell}(x)$ is the Legendre polynomial of order $\ell$. $P_{\rm NL}(k,\chi)$ is the 3D non-linear matter power spectrum at wavenumber k (which, in the Limber approximation, is set equal to $(l + 1/2)/ \chi(z_i)$) and at the cosmic time associated with redshift $z_i$. We find that the redshift evolution of ${w}_{\rm DM}(z_i)$ depends little on the particular value of cosmological parameters, whereas the dependence of the overall amplitude of ${w}_{\rm DM}(z_i)$ with respect to cosmology is absorbed by our systematic functions. Based on this, we hold cosmology fixed when computing ${w}_{\rm DM}(z_i)$, assuming the values in \citealt{Planckresults2018}). We then verify \textit{a posteriori} that this approximation is valid by repeating our analysis using very different values for the cosmological parameters ($\Omega_{\rm m} = 0.4$, $\sigma_8=0.7$), finding that the impact on our conclusions is negligible. Note that some of the mitigation schemes adopted in literature to correct the galaxy-matter bias evolution of the unknown sample also automatically estimate ${w}_{\rm DM}(z_i)$ from the data \citep{MattewsNewman2010,vandenBusch2020}, but they are not adopted in this work.

 \item \textbf{Magnification \gbso{effects}{signal} ${M}(z_i)$}. Weak lensing magnification \citep{Narayan1989,Villumsen1997,Moessner1998} \gbso{can lead to a change in} changes the observed spatial density of galaxies: the enhancement in the flux of magnified galaxies can locally increase the number density, as more galaxies pass the selection cuts/detection threshold of the sample; at the same time, the same volume of space appears to cover a different solid angle on the sky, generally causing the observed number density to decrease. The net effect is driven by the slope of the luminosity function, and it has an impact on the measured clustering signal. Considering only the dominant terms \giulia{(which account for the magnification of the unknown sample by the reference sample and the magnification of the reference sample by the unknown sample}) {and assuming linear bias,} this can be written as:
    
\begin{multline}
\label{eq:magnification}
        {M}(z_i) = \int d\theta W(\theta) \int \frac{dl \ l}{2\pi} J_0(l\theta)\int  \frac{d\chi}{\chi^2} \\
        \times\left[b_{\rm r} \alpha_{\rm u} q^{\rm r}_{\delta}q^{\rm u}_{\kappa} 
        + b_{\rm u} \alpha_{\rm r} q^{\rm u}_{\delta} q^{\rm r}_{\kappa}\right]P_{\rm NL}\left(\frac{l+1/2}{\chi},z(\chi)\right),
\end{multline}
where the terms $q_{\delta}$ and $q_{\kappa}$ read:
\begin{equation}
q_{\delta}(\chi) = n[z(\chi')]\frac{dz}{d\chi'},
\end{equation}
\begin{equation}
q_{\kappa}(\chi) = \frac{3H_0^2{\Omega}_{\rm m}\chi}{c^2a(\chi)} \int_{\chi}^{\chi(z=\infty)} d\chi' {n(z(\chi'))}\frac{dz}{d\chi'} \frac{\chi'-\chi}{\chi'}.
\end{equation}
In the above equations, $n[z(\chi)]$ is either $n_{\rm u}(z)$ or $n_{\rm r,i}(z)$
Under the approximation of thin redshift bins, we can \marco{write} Eq.~\ref{eq:magnification} as a discrete summation over redshift bins of width $\Delta \chi$:
%
\begin{multline}
\label{eq:linear_mag}
{M}(z_i) =  b_{\rm r}(z_i) \alpha_{\rm u}(z_i)  \sum_{j>i} \left[D_{ij} n_{\rm u}(z_j)\right]+ b_{\rm u}(z_i) \alpha_{\rm r}(z_i)  \sum_{j>i} \left[ D_{ij}n_{\rm u}(z_i )\right],
\end{multline}
with
\begin{equation}
D_{ij}  = \frac{3H_0^2{\Omega}_{\rm m}}{c^2} {w}_{\rm DM}(z_i)\frac{\chi(z_i)}{a(z_i)} \frac{\chi(z_j)-\chi(z_i)}{\chi(z_j)} \Delta \chi_j.
\end{equation}

The magnification coefficient $\alpha\equiv 2.5 s-1$ is related to the slope  $s$ of the cumulative number counts evaluated at flux limit. The slope of the cumulative number counts is formally defined for a flux limited sample as
\begin{equation}
\label{eq:slope}
s= \frac{d}{dm} \log_{10} n(<m),
\end{equation}
where $n(<m)$ is the cumulative number count as a function of magnitude $m$, and $s$ is to be evaluated at the flux limit of the sample. For a sample which is not flux limited, evaluating the coefficient $s$ is more complicated, \marco{and Eq.~\ref{eq:slope} can not be used.} \marco{Estimates of $\alpha$ for both the reference and unknown samples are needed to properly model magnification effects.}\
\end{itemize}

Under the assumption of thin reference bins, linear galaxy-matter bias, and using the linearised version of the equation describing magnification effects (Eq.~\ref{eq:linear_mag}) , Eq.~\ref{menard2} becomes a linear system of equations:

\begin{multline}
\label{system}
{w}_{\rm ur}(z_i) =  n_{\rm u}(z_i) b_{\rm u}(z_i) b_{\rm r}(z_i)
 {w}_{\rm DM}(z_i) + \\b_{\rm r}(z_i) \alpha_{\rm u}(z_i)  \sum_{j>i} \left[D_{ij} n_{\rm u}(z_j)\right]+ b_{\rm u}(z_i) \alpha_{\rm r}(z_i)  \sum_{j>i} \left[D_{ij} n_{\rm u}(z_i ) \right].
\end{multline}
If the  values $b_*(z_i)$, $\alpha_*(z_i)$, ${w}_{\rm DM}(z_i)$ are provided, this can be solved to obtain an estimate of $n_{\rm u}(z_i)$. This would be similar to standard clustering-based methods which use the cross-correlation signal as a starting point to infer the redshift distributions of the unknown sample  \citep{Newman2008,Menard2013,Schmidt2013,McQuinn2013}. 
Alternatively, if an estimate of the $n_{\rm u}(z_i)$ is provided by, e.g., a photo-$z$ method, Eq.~\ref{system} can be used to \gbso{quickly} evaluate the expected correlation signal ${w}_{\rm ur}(z_i)$ and compare it to the one measured in data, {i.e.} \gbso{This can be interpreted as} a forward modelling approach (see, e.g., \citealt{Choi2016}).

\gbso{We note that} This work represents a significant advancement over DES Y1, because in the Y1 analysis none of the terms described above were modeled. We assumed $b_{\rm r}(z_i)$, $b_{\rm u}(z_i)$, ${w}_{\rm DM}(z_i)$ to be constant within each photo-$z$ bin, and used the simulations to estimate the systematic error induced by this assumption. In DES Y1 we also did not model $M(z_i)$, but we decided to exclude the redshift range (i.e., the tails of the redshift distributions) where magnification effects are expected to have a non-negligible impact. On the contrary in this work we model $b_{\rm r}(z_i)$, ${w}_{\rm DM}(z_i)$ and, depending on the method, $M(z_i)$.

\subsection{Assigning likelihood to the cross-correlation information}
\label{sec:combination}
\gbso{In DES Y3} We use the clustering data {$\{w_{\rm ur}(z_i), w_{\rm rr}(z_i)\},$ generically referred to as WZ, to place a likelihood $\mathcal{L}\left[WZ | n_{\rm u}(z)\right]$ of obtaining the WZ data given some estimate of the true $n_{\rm u}(z).$ The WZ data will be used to evaluate the likelihood of many candidate $n_{\rm u}(z)$ functions, typically drawn from some combination of PZ and spectroscopic data.  In the DES Y1 analysis, such realisations were taken as $n_{\rm u}(z) = n_{\rm pz}(z+\Delta z),$ where $n_{\rm pz}(z)$ was a single ``best'' photo-$z$ estimate and $\Delta z$ a free parameter.  The Y3 approach is more general, with many realizations of the full function $n_{\rm u}(z)$ being drawn.  In any case we need only to define $\mathcal{L}\left[WZ | n_{\rm u}(z)\right].$  To do so, we make use of two approaches, described below.}

\subsubsection{Mean-matching method}
{This method works by compressing the $n(z)$ functions to a single statistic, their mean $\avg{z}.$ \marco{In this ``simpler'' method, we do not model magnification effects}, so the mean is taken over a restricted range of $z$, where a reference sample is available and $w_{\rm ur}(z)\gg M(z),$} such that we can neglect magnification effects.  \marco{For this method, cutting the tails can be preferable even when estimates of magnification effects in the tails are available. This is due to the fact that small errors in the magnification estimates in the tails can have a large impact on the mean of the redshift distribution, lowering the capability of the method to constrain the mean redshift.}
 
 Following \gbso{what was done in} the DES Y1 analysis, we choose {a fixed} interval $[z_{\rm min},z_{\rm max}] =[\avg{z}_{\rm pz}-2\sigma_{\rm pz},\avg{z}_{\rm pz}+2\sigma_{\rm pz}]$, where {$\avg{z}_{\rm pz}$ and} $\sigma_{\rm pz}$ {are the mean and} root mean square of {a canonical $n_{\rm pz}(z)$}.  \giulia{In case the fixed interval includes a range where there is no reference sample coverage, it is further reduced to ensure there are enough galaxies in the reference sample to provide a meaningful WZ estimate} (see Section~\ref{sect:systematic_method1} for more details). 
\gbso{Starting from the cross-correlation measurements ${w}_{\rm ur}(z_i)$, we estimate the redshift distribution using Eq.~\ref{system}:} {We first create a nominal {``naive''} estimator $\tilde n_{\rm u}(z)$ using Eq.~\ref{system} which would be proportional to an unbiased estimator if linear bias holds and $b_{\rm u}(z)$ is constant:}
    %
    \begin{equation}
    \label{eq:final_estimator}
\tilde n_{\rm u}(z_i) \propto \frac{{w}_{\rm ur}(z_i)}{ b_{\rm r}(z_i) {w}_{\rm DM}(z_i)},
    \end{equation}
    {Then we define mean redshifts for the WZ data and the proposed $n_{\rm pz}(z)$  as}
\begin{align}
\label{eq:windowed_mean}
\avg{z}_{\rm wz} & = \frac{\int_{z_{\rm min}}^{z_{\rm max}} dz\, z\, \tilde n_{\rm u}(z)}{\int_{z_{\rm min}}^{z_{\rm max}} dz\, \tilde n_{\rm u}(z)} \\
 \avg{z}_{\rm pz} & = \frac{\int_{z_{\rm min}}^{z_{\rm max}} dz\, z\, n_{\rm pz}(z)}{\int_{z_{\rm min}}^{z_{\rm max}} dz\,  n_{\rm pz}(z)}
\end{align}
      
\gbso{where $A$ is a normalisation constant which has no relevant effect in this method. In the above equation the term $b_{\rm u}(z_i)$ does not appear as we cannot estimate it, and its absence must be compensated by an appropriate systematic uncertainty term. We then compare the mean of the clustering-based redshift distribution estimate with the mean of the distributions $\{n^{{\rm pz},k}_{\rm u}(z_i)\}$ in the chosen interval. This is similar to what was done in DES Y1, where we used clustering-based estimates to directly correct the mean of the posterior of a given photo-$z$ code; the main difference is that now we further correct for the terms $b_{\rm r}(z_i)$,  ${w}_{\rm DM}(z_i)$ in the clustering-based estimate.}
    
The likelihood of the WZ data given a proposed $n_{\rm u}(z)$ is then taken to be a Gaussian distribution in the $\avg{z}$ values:
\begin{equation}
  \label{likeY1}
  \mathcal{L}\left[ \textrm{WZ} | n_{\rm u}(z)\right] \equiv \mathcal{N}\left( \avg{z}_{\rm pz}-\avg{z}_{\rm wz}, \sigma_{\avg{z}}\right)
    \end{equation}
    \gbso{In the above equation, $\avg{z}_{{\rm pz},k}$ and $ \avg{z}_{{\rm wz}}$ are the mean of the redshift distributions estimated by the $k$-th photo-$z$ realisation and by the clustering-based method. The prior on the mean in Eq.~\ref{likeY1} takes into account systematic uncertainties of the methods. The quantity $\hat{\Sigma}^{-1}_{\avg{z}_{\rm wz}}$ is the precision matrix of the clustering-based mean redshift measurement.}
{The uncertainty $\sigma_{\avg{z}}$ must incorporate the estimated measurement noise and also systematic errors from shortcomings of the underlying model.  Section~\ref{sect:systematic_method1} gives the results of using simulations to set these uncertainties. The assumption of Gaussianity is a reasonable choice even in absence of systematics, as per the central limit theorem (the mean redshift compresses the information from many different redshifts). Moreover, we parametrise the impact of systematics effects in such a way they can be described by a Gaussian likelihood, and systematic effects dominate our total error budget. }

    
    
\subsubsection{Full-shape method}\label{sect:fsmethod}
{This method dispenses with the mean statistic and simply compares the observed $w_{\rm ur}(z_i)$ data to a model $\hat w_{\rm ur}[z_i; n_{\rm u}(z), b_{\rm r}(z), \alpha_{\rm r}(z), \vecs, \vecp]$ that incorporates potential systematic effects.  The model is an alteration of Eq.~\ref{system}:}
\begin{multline}
\label{wursys}
{\hat w}_{\rm ur}(z_i) =  n_{\rm u}(z_i) b_{\rm r}(z_i)
 {w}_{\rm DM}(z_i) \times \sys(z_i,\vecs) +\\
b_{\rm r}(z_i) \alpha_{\rm u}^{'}(z_i)  \sum_{j>i} \left[D_{ij} n_{\rm u}(z_j)\right]+ b_{\rm u}^{'}(z_i) \alpha_{\rm r}(z_i)  \sum_{j>i} \left[D_{ij} n_{\rm u}(z_j ) \right].
\end{multline}
{
The functions $n_{\rm u}(z), b_{\rm r}(z),$ and $\alpha_{\rm r}(z)$ are assumed to be given beforehand, and $w_{\rm DM}$ is calculated from theory as described in Eq.~\ref{eq:clustzonly}.  The $\sys$ function multiplies the clustering signal by some redshift-dependent value that is parameterized by $\vecs=\{s_1,s_2,\ldots\}$ that we will marginalize over.   The role of the $\sys$ function is to absorb all uncertainties in $b_{\rm u}$ and its redshift dependence, as well as uncertainties due to failures in the linear bias model itself, and in the determination of $b_{\rm r}(z).$  The choice of $\sys$ function and the priors on its parameters are guided by simulations as described in \S\ref{sect:systematic_method2}.  The magnification terms in the second line of Eq.~\ref{wursys} are scaled by the unknowns $\vecp=\{b_{\rm u}^{'}, \alpha_{\rm u}^{'}\}.$ Leaving these as free parameters absorbs uncertainties not only in these values but also in $b_{\rm r},\alpha_{\rm r}$ and in the linear-bias model adopted for magnification.  For this reason the $b_{\rm u}^{'}$ value appearing in the magnification is not assumed to equal the $b_{\rm u}$ that might multiply $w_{\rm DM}.$  We do not implement redshift dependence of $\vecp$ (although the formalism would allow it) because magnification signals are important only over limited ranges of $z$ (i.e., in the tails, see, e.g., \citealt*{Gatti2018}) for a given tomographic bin of the WL sources.}

\gbso{we forward model the cross-correlation signal ${\hat{w}_{\rm ur}}(n^{\rm pz,k})$ using Eq.~\ref{system} and using as $n_{\rm u}(z_i)$ the $k$-th realisation provided by a given photo-$z$ code. Then, we compute the likelihood with the measured cross-correlation signal ${{w}_{\rm ur}}(z_i)$ in data. In this case, the likelihood can be written as:}

{With a model for $w_{\rm ur}$ in hand, we assume that the measurement errors in the data are Gaussian and define a likelihood}
%
    %
\begin{multline}
\label{eq:wzlike}
  \mathcal{L}\left[{\rm WZ} | n_{\rm u}(z), b_{\rm r}(z), \alpha_{\rm r}(z), w_{\rm DM}(z)\right]
  \propto \\
  \int d\vecs\, d\vecp\, \exp\left[ -\frac{1}{2} (w_{\rm ur}-\hat w_{\rm ur})^T \Sigma_w^{-1} (w_{\rm ur}-\hat w_{\rm ur}) \right] p(\vecs) p(\vecp).
\end{multline}

{The data and model for $w_{\rm ur}$ are taken here to be vectors over $z_i$, and $\Sigma_w$ is the covariance matrix of the data (from shot noise and sample variance).  The nuisance parameter sets \vecs\ and \vecp\ each have their own priors.  It is the extent of these priors that regulates the level of systematic error allowed for in the inference of $n_{\rm u}(z)$ from the WZ data.
The systematic function and these priors are quantified in \S\ref{sect:systematic_method2}. }
  
\gbso{The model $\hat{w}_{\rm ur}$ is described by Eq.~\ref{system} multiplied by a function ${\rm Sys (z,\{s_i\})}$ that accounts for the systematic uncertainties of the method. The nuisance parameters $\{s_i\}$ are the parameters of the systematic functions; on the other hand, $\{ p_i\}$ are the free parameters that appear in Eq.~\ref{system}, i.e., the magnification parameters $\alpha_{\rm u}$,$\alpha_{\rm r}$, and the bias of the unknown sample $b_{\rm u}(z)$.}

The covariance matrix $\Sigma_w$ \gbso{for each of the likelihoods is the appropriate covariance matrix from the cross-correlation analysis. They are} is estimated from simulated data through a jackknife (JK) approach, using the following expression \citep{Quenouille1949,Norberg2009}:
\begin{equation}
\hat{\Sigma}(x_i,x_j)=\frac{(N_{\rm JK}-1)}{N_{\rm JK}} \sum_{k=1}^{N_{\rm JK}} (x_i^k-\bar{x_i})(x_j^k-\bar{x_j}),
\end{equation}
where the sample is divided into $N_{\rm JK}=1000
$ sub-regions of roughly equal area,  $x_i$ is a measure of the statistic of interest ($=w_{\rm ur}$) in the $i$-th bin of the $k$-th sample, and $\bar{x_i}$ is the mean of the resamplings. The jackknife regions are safely larger than the maximum scale considered in our clustering analysis. The Hartlap correction \citep{Hartlap2007} is used when computing the inverse covariance.

{
Note that the WZ likelihood in Eq.~\ref{eq:wzlike} depends explicitly on the estimated bias and magnification coefficient $b_{\rm r}$ and $\alpha_{\rm r}$ of the reference sample, and depends implicitly on the cosmological model through the dark-matter clustering $w_{\rm DM}.$  Thus in principle, this likelihood and the inferences on $n_{\rm u}(z)$ must be recalculated for each change in cosmological model.  We have, however, tested numerically that the full expression for $\mathcal{L}[\textrm{WZ}|n(z)]$ has negligible dependence on the cosmological parameters or the reference-sample properties once the marginalization over systematic nuisances \vecs\ and \vecp\ are done.  This is because the systematic variables have enough freedom to absorb the small changes in the model wrought by changes in cosmology.  It is therefore allowable for us to compute Eq.~\ref{eq:wzlike} using a fiducial cosmology and fiducial values of $b_{\rm r},\alpha_{\rm r},$ and use the inferred redshift distributions in a cosmological inference that might vary these parameters.}

\section{Data and simulated data}\label{sect:data}
{
This section describes the various photometric and spectroscopic catalogs that feed into the clustering-$z$ measurements.  The full analysis is also conducted on simulated catalogs; for each element of the real analysis, we also describe how its simulated counterpart was generated.}

\subsection{DES Y3 data}
The Dark Energy Survey (DES) observed $\sim$ 5000 square degrees of the southern hemisphere in 5 different broad photometric bands ($grizY$) over \gbso{the past} six years\gbso{DES  has been collecting images with} {using} the Dark Energy Camera (DECam, \citealt{Flaugher2015}), a 570-megapixel camera built by the DES Collaboration and stationed at the Cerro Tololo Inter-American Observatory (CTIO) 4-meter Blanco telescope. \gbso{and it is expected to infer} {DES will measure} the shapes of about 300 million galaxies up to redshift $z \sim 1.4$. In this paper we focus on the \gbso{DES Year 3 (Y3) data, which comprises the} analysis of the first three years (Y3) of observations. DES Y3 data span the full area of the survey,  4143 deg$^2$ after masking for foregrounds and problematic regions, a {major advance} over the 1321 deg$^2$ of DES Y1 \gbso{area ($\sim$ used in the Y1 cosmic shear analysis} (\citealt{Troxel2018,y1gold}). \gbso{but it does not reach yet the maximum depth.} {The complete DES (Y6) reaches greater depth than Y3 data; furthermore, the data is more uniform in depth.}. The total number of objects detected in DES Y3 is $\approx390,000,000.$ Object detection and measurements are described in \cite{y3-gold}.

\subsection{Buzzard N-body simulation}\label{sect:sims_buzzard}
We use one realisation of the DES Y3 Buzzard catalog v2.0 \citep{DeRose2018}. Initial conditions were generated using 2LPTIC \citep{Crocce2006} and the N-body run using L-GADGET2 \citep{springel2005}. Cosmological parameters have been chosen to be  $\Omega_{\rm m} = 0.286$, $\sigma_8 = 0.82$, $\Omega_b = 0.047$, $n_s = 0.96 $, $h = 0.7$. Lightcones are generated on the fly starting from three boxes with different resolutions and size (1050$^3$, 2600$^3$ and 4000$^3$ Mpc$^3 h^{-3}$ boxes and 1400$^3$, 2048$^3$ and 2048$^3$ particles), to accommodate the need of a larger box at high redshift. Halos are identified using the public code ROCKSTAR \citep{Behroozi2013} and they are populated with galaxies using ADDGALS \citep{DeRose2018}. Galaxies are assigned magnitudes and positions based on the relation between redshift, $r$-band absolute magnitude and large-scale density found in a subhalo abundance matching model \citep{Conroy2006,Lehmann2017} in higher resolution N-body simulations. SEDs are assigned to galaxies from the SDSS DR7 {Value Added Galaxy Catalog} \citep{Blanton2005} by imposing the matching with the SED-luminosity-density relationship measured in the SDSS data. SEDs are $K$-corrected and integrated over the DES filter bands to generate DES $grizY$ magnitudes.
Lensing effects are calculated using the multiple plane ray-tracing algorithm CALCLENS \citep{Becker2013}, which provides weak-lensing shear, magnification and lensed galaxy positions for the lightcone outputs. \gbso{CALCLENS is run onto the sphere using the HEALPix algorithm and is accurate to $\sim 6.4$ arcseconds.}

\subsection{Reference sample 1: \redmagic\ galaxies}
\begin{figure}
\begin{center}
\includegraphics[width=0.5 \textwidth]{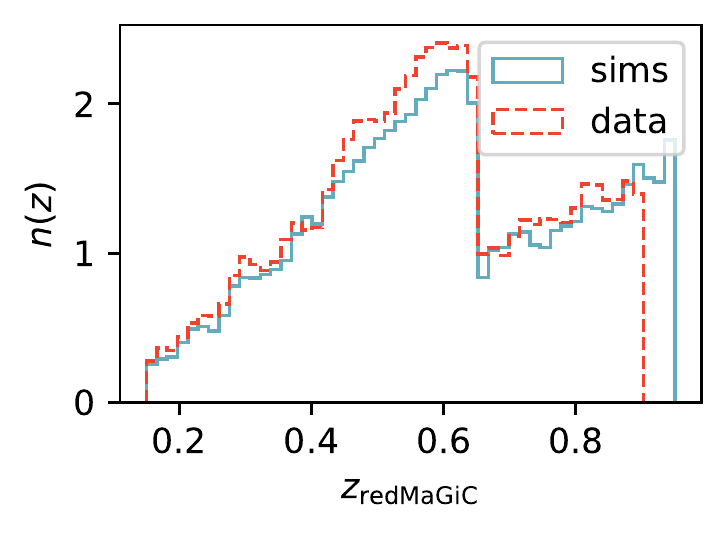}
\end{center}
\caption{Redshift distributions of the \redmagic\ samples, binned using the \redmagic\ photo-$z$ estimates, in data and in simulations.} 
\label{fig:Nz_redmagic_sims_data}
\end{figure}
\begin{figure*}
\begin{center}
\includegraphics[width=0.95 \textwidth]{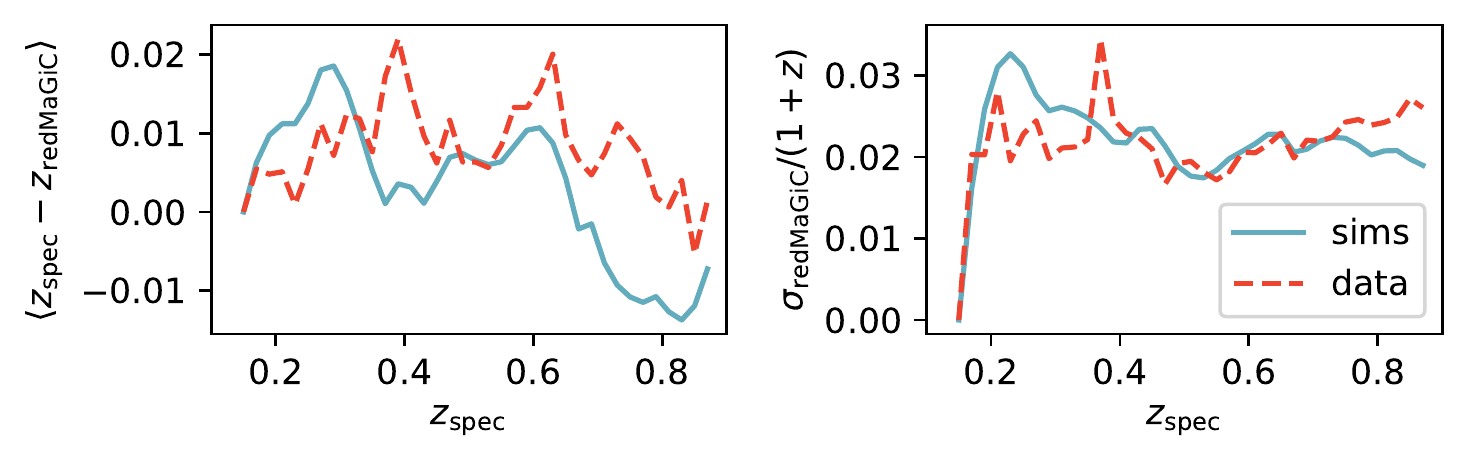}
\end{center}
\caption{The bias (left) and scatter (right) of $z_{\redmagic}$ for the simulated \redmagic\ sample (solid lines) compared to the data (dashed lines).} 
\label{fig:redmagic_unc}
\end{figure*}

The first reference sample used in this WZ analysis consists of DES \redmagic\ galaxies. The \redmagic\ algorithm selects red luminous galaxies with high quality photometric redshift estimates \citep{Rozo2016}. This is achieved by fitting each galaxy to a red sequence template; galaxies are then selected only if they pass a goodness of fit and luminosity threshold. In DES, \redmagic galaxies are used as lens sample in the galaxy-galaxy lensing analysis and in the clustering analysis \citep{y3-gglensing,y3-galaxyclustering}. Two samples are selected with different number density by means of two distinct luminosity thresholds: a first sample called ``high density'' selected with a cut L/L* > 0.5 and a sample called ``high luminosity'' selected with a cut L/L* > 1. A combined sample is then obtained by joining these two samples, using the high density sample for redshifts $z<0.65$, the high luminosity sample for higher redshifts.

In simulations, the \redmagic\ sample is selected with the same algorithm used in the data. A comparison between the redshift distributions for the \redmagic\ samples in data and in simulations is shown in Fig.~\ref{fig:Nz_redmagic_sims_data}, illustrating the good agreement between the two. Small differences are due to
small discrepancies in the evolution of the red-sequence between the simulation and the data. Both in simulations and in data, the \redmagic\ sample is divided into {40 bins of width $\Delta z=0.02$ spanning the $0.14<z<0.94$ range of the \redmagic\ catalog.}\footnote{We note that the simulated \redmagic\ sample spans a slightly wider range in redshift; we nonetheless cut the redshift interval at $z$=0.90 to be consistent with the data.}. The particular choice of the bin width is not expected to impact our conclusions, as long as bins are small enough compared to the typical variation scales of the the weak lensing $n(z)$ and the galaxy-matter biases of the two samples. The total number of \redmagic\ galaxies is 3,041,935 in the data, and  2,594,036 in the simulation. This implies that the statistical uncertainties of the clustering redshift estimates obtained using the \redmagic\ sample are larger in simulations compared to data. We do not expect this to be important, as we show in Section~\ref{sect:systematic_method1} that the clustering-$z$ methodology is dominated by systematic uncertainties, and the statistical uncertainties are negligible.


We compare the typical \redmagic\  photo-$z$ scatter  and  bias found in data vs in simulations in Fig.~\ref{fig:redmagic_unc}. Since only a portion of the data have spec-z information, we re-weight the magnitude distribution of the spectroscopic sample such that it matches the magnitude distribution of the \redmagic\ galaxies before computing the statistics shown in Fig.~\ref{fig:redmagic_unc}. This re-weighting is performed separately for each redshift bin. \marco{Note that the typical scatter of \redmagic\ photo-$z$ is similar to our bin width, which might call into question the choice of bin width for \redmagic\ galaxies. However, we verify in Section~\ref{sect:systematic_method1} that even with this setup, \redmagic\ photo-$z$ uncertainties are not a dominant source of systematic error for our methodology. Therefore, we decided using a larger bin width for \redmagic\ galaxies was not necessary. }

\marco{Using cross-correlation techniques, \cite{y3-lenswz} noted that photo-$z$ uncertainties in \redmagic\ galaxies at $z>0.8$ might be underestimated. We do not think this constitutes a problem for the current analysis, as \redmagic\ photo-$z$ uncertainties are a sub-dominant systematic in our methodology (Section~\ref{sect:systematic_method1}), and WZ constraints at $z>0.8$ are driven by the BOSS/eBOSS sample (Section~\ref{ss:ss}).}

A catalog of random points for \redmagic\ galaxies is generated uniformly over the footprint. Both in data and in simulations,  weights are assigned to \redmagic\ galaxies such that spurious correlations with observational systematics are cancelled. \marco{Note that due to low-statistics issues, the weights do not resolve fluctuations on scales relevant for this work, but only captures large-scale spurious correlations.} The methodology used to assign weights is described in \cite{y3-galaxyclustering}, and it is the same for data and simulations. The main difference between data and Buzzard simulations is that the latter only models depth variations across the footprint, while data are subject to a larger number of systematics which are not modelled in simulations. This should not affect any conclusion drawn here as the weights effectively remove all the spurious dependence of the number density with respect to any systematic, regardless of their number.


\marco{Finally, estimated magnification coefficients for \redmagic\ galaxies are obtained using the Balrog image simulations \citep{Suchyta2016,y3-balrog} in a process briefly described here. Galaxy profiles are drawn from the DES deep fields \citep*{y3-deepfields} and injected into real DES images. The full photometry pipeline \citep{y3-gold} and \redmagic\ sample selection are applied to the new images to produce a simulated \redmagic\ sample with the same selection effects as the real data. To compute the impact of magnification, the process is repeated, this time applying a constant magnification to each injected galaxy. The magnification coefficients are then derived from the fractional increase in number density when magnification is applied. This method captures the impact of magnification on both the galaxy magnitudes and the galaxy sizes, including all sample selection effects. See \cite{y3-balrog,y3-2x2ptmagnification} for further details. The coefficients have been estimated for \redmagic\ in 5 wide redshift bins, centered at $z = (0.25,  0.425, 0.575, 0.75,  0.9)$: yielding \giulia{the magnification coefficients} $\alpha_{\rm r} = (0.313 , -1.515 , -0.6628,  1.251 ,  0.9685)$. We interpolate these values in $z$  using the \texttt{scipy} routine \texttt{interp1d} to provide an estimate of $\alpha_{\rm r}$ for every reference bin used in the clustering analysis.  }

\subsection{Reference sample 2: spectroscopic galaxies}
\begin{figure}
\begin{center}
\includegraphics[width=0.5 \textwidth]{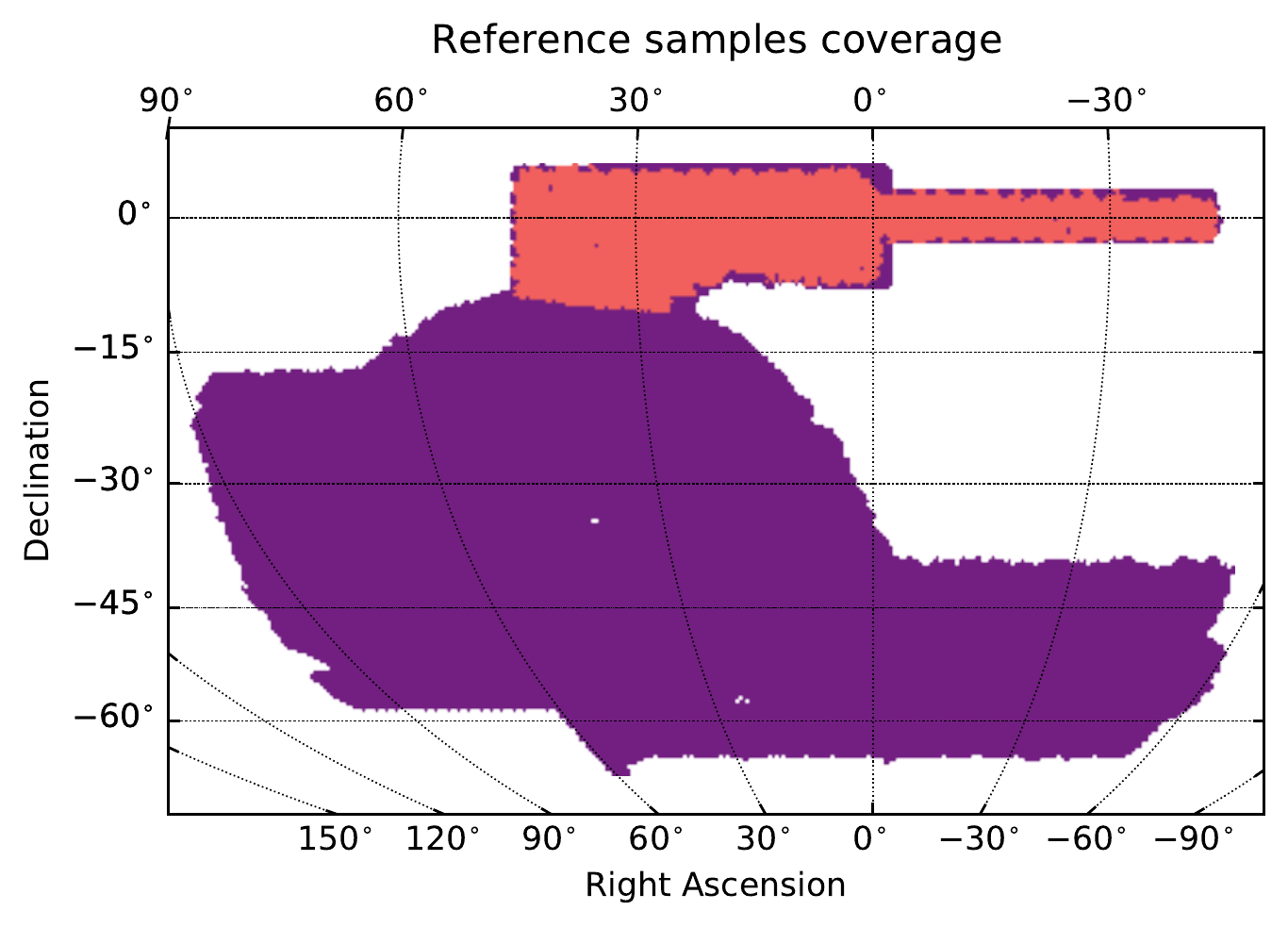}
\end{center}
\caption{Spatial coverage of the two reference samples used in this work. Purple indicates the coverage by \redmagic\ galaxies, pink indicates he coverage by BOSS and eBOSS galaxies.}
\label{fig:coverage}
\end{figure}

\begin{figure}
\begin{center}
\includegraphics[width=0.45 \textwidth]{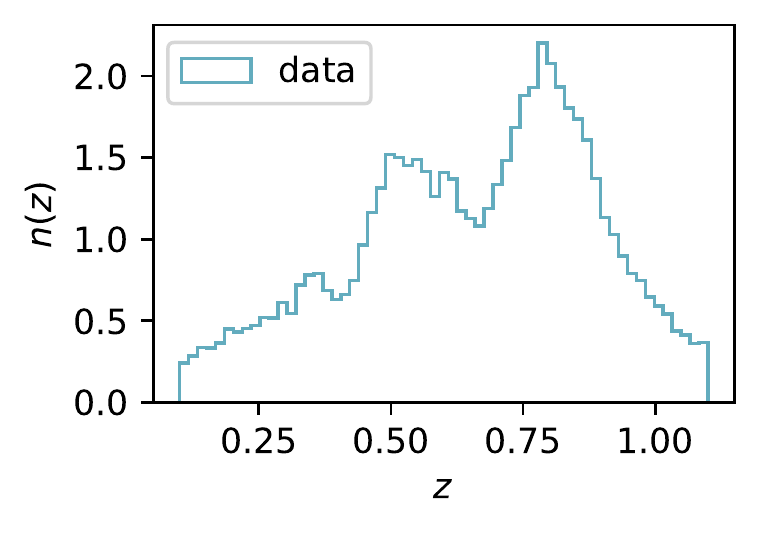}
\end{center}
\caption{Redshift distribution of the BOSS/eBOSS sample in data and in simulations.}
\label{fig:ebossnz}
\end{figure}

The second reference sample used in this work is a combination of spectroscopic samples from the Sloan Digital Sky Survey (SDSS, \citealt{Gunn2006,Eisenstein2011,Blanton2017}). In particular,  we combine SDSS galaxies from BOSS (Baryonic Oscillation Spectroscopic Survey, \citealt{Smee2013,boss}) and from eBOSS (extended-Baryon Oscillation Spectroscopic Survey, \citealt{eboss2016,dr16,Alam2020}). The BOSS sample includes the LOWZ and CMASS catalogs from the SDSS DR 12, fully described in \cite{reid16}, while we included the large-scale structure catalogs from emission line galaxies (ELGs, see \citealt{raichoor17} for the target selection  description), luminous red galaxies (LRGs, target selection described in \citealt{prakash16}) and quasi stellar objects (QSOs) (eBOSS in prep.) from eBOSS, which were provided to DES for clustering redshifts usage by agreement between DES and eBOSS. \marco{The different samples are stacked together, and used as one single reference sample in this work. Each sample comes with its own catalog of random points, which account for selection effects. Different catalogs of random points are stacked together. We made sure the ratio of the number of randoms with respect to the number of galaxies was the same for each random catalog before combining them.} \giulia{Both in simulations and in data, the BOSS/eBOSS sample is divided into 50 bins spanning the $0.1<z<1.1$ range of the catalog (width $\Delta z\sim0.02$). } The redshift distribution of the samples is shown in Fig.~\ref{fig:ebossnz} and the area coverage and number of objects of each sample is summarised in Table~\ref{table:spec_ngals}. The area coverage is smaller compared to \redmagic\ galaxies, as shown in Fig.~\ref{fig:coverage}. \marco{Note that some of the galaxies in the BOSS/eBOSS sample are also in the \redmagic\ catalog: $\sim$1 per cent of the \redmagic\ galaxies are matched to $\sim$10 per cent of the BOSS/eBOSS galaxies, within 1 arcsec. We did not remove these galaxies from the \redmagic\ sample, as they have a negligible impact both on our constraints and on the covariance between the two samples (as it will be clear in the following sections, the constraints from both samples are systematic-dominated). }

To replicate the spectroscopic BOSS/eBOSS sample in simulations, we selected bright galaxies with similar sky coverage and redshift distribution as the ones in data.  We did not try to further match other properties of the sample, {e.g. the galaxy-matter bias likely differs from that of the real data.  We note that the WZ methodology corrects for the reference bias, so at no point in the analysis of the real data are we assuming that the simulations have the same bias.} \gbso{We note that even if the galaxy-matter bias of the BOSS/eBOSS sample selected in simulation might differ from the one measured in data, all the methods implemented in this paper will correct for this.}

\marco{Last, we note that estimates of the magnification coefficients are not available for BOSS/eBOSS galaxies, as we did not try to reproduce the complex BOSS/eBOSS selection function within Balrog image simulations.} This is not a problem, as we verify in this work that our analysis is not very sensitive to the particular choice of the values of the magnification parameters. For our fiducial analysis, we assumed magnification values for the BOSS/eBOSS sample similar to the \redmagic\ ones.

\newpage

\begin{table}
\begin{center}
    \begin{tabular}{|c|c|c|c|}
      \hline
      \multicolumn{4}{|c|}{Spectroscopic Samples} \\
      \hline
      Name & Redshifts & $N_{\rm gal }$ & Area \\
      \hline
      LOWZ (BOSS) & $z \sim [0.0,0.5]$ & $45671$ & $\sim860 \  \text{deg}^2$   \\   
      \hline
      CMASS (BOSS) & $z \sim [0.35,0.8]$ & $74186$ & $\sim860 \  \text{deg}^2$  \\ 
      \hline
      LRG (eBOSS) & $z \in [0.6,1.0]$ & $24404$ & $\sim700 \  \text{deg}^2$  \\   
      \hline
      ELG (eBOSS) & $z \in [0.6,1.1]$ & $89967$ & $\sim620 \  \text{deg}^2$  \\   
      \hline
      QSO (eBOSS) & $z \in [0.8,1.1]$ & $7759$ & $\sim700 \  \text{deg}^2$  \\   
      \hline
    \end{tabular}
  \caption{List of the spectroscopic samples from BOSS/eBOSS overlapping with the DES Y3 footprint used as reference galaxies for clustering redshifts in this work.}
  \label{table:spec_ngals}
  \end{center}
\end{table}

\subsection{Weak lensing sample}\label{sect:data_wl}

\begin{figure*}
\includegraphics[width=0.95\textwidth]{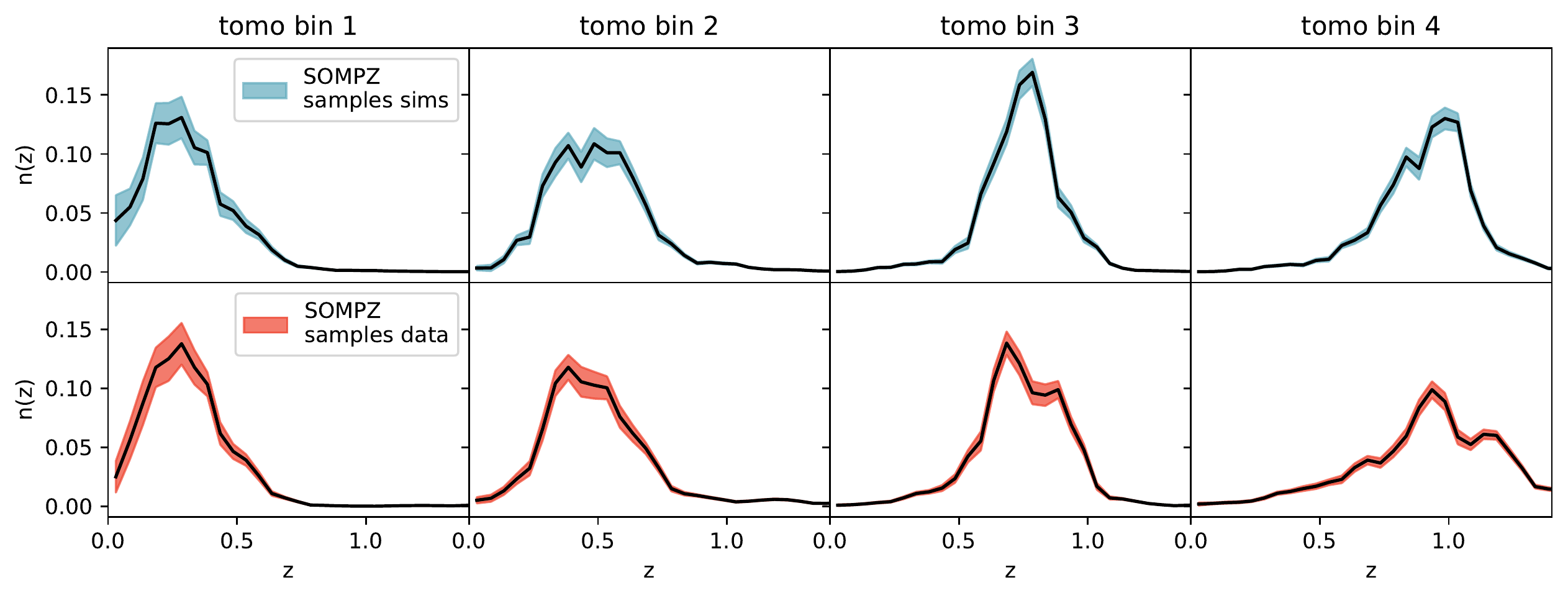}
\caption{SOMPZ redshift distributions, as estimated in simulations (upper panels) and in data (lower panels), for the four tomographic bins considered in this analysis. The bands represent the 68\% confidence interval spanned by the SOMPZ n(z) realisations.}
\label{fig:Nz_sims_data}
\end{figure*}

The weak lensing sample in data is created using the \mcal\ pipeline, which is fully described in \cite*{y3-shapecatalog}. After creation of the DES Y3 `Gold` catalog \citep{y3-gold}, the \mcal\ pipeline measures the shapes of each detected object. \gbso{The \mcal\ pipeline is able to self-calibrate the measured shapes against shear and selection biases by measuring the mean shear and selection response matrix of the sample $\langle \mcalR \rangle = \langle \mcalRg \rangle +  \langle \mcalRs \rangle$. The current DES Y3 implementation of \mcal\ is able to correct for shear biases up to a multiplicative factor of 2-3 per cent, which is fully characterised using image simulations}
Selection cuts for the sample are described in \cite*{y3-shapecatalog} and are {chosen from results of tests on both sky data and} \gbso{empirical tests and tests on} image simulations  \citep{MacCrannSims2019}, and are designed to minimise systematic biases in the shear measurement. Galaxies are {weighted by the inverse variance of shear measurement,} which increases the statistical power of the catalog. The final sample comprises $\approx100$~million objects, for an effective number density of $n_{\rm eff} = 5.59$ gal/arcmin$^{-2}$. Galaxies are further divided into 4 tomographic bins, and redshift {distribution} estimates for each of the tomographic bins are provided by the SOMPZ method (\citealt{Buchs2019}; \citealt*{y3-sompz}). The tomographic bins are selected such that they have {roughly} equal raw number density. \gbso{through an iterative procedure.}

The weak lensing sample is reproduced with high fidelity in the Buzzard simulation by applying flux and size cuts to the simulated galaxies that mimic the DES Y3 source selection thresholds.
\gbso{The WL galaxy sample in Buzzard is selected with the aim of reproducing the same selection applied in DES Y3 data in terms of size, signal-to-noise and colors.} Shape noise has been added to the galaxies to match the measured shape noise of the DES Y3 WL sample. \marco{Estimates of the magnification coefficients are obtained using Balrog image simulations \citep{y3-balrog}, similarly to \redmagic\ galaxies.  In particular, for the four tomographic bins, we adopt the values $\alpha_{\rm u} = (-0.443, -0.209, -0.004,  0.306)$ \citep{y3-2x2ptmagnification}}. \marco{These values do not need to be interpolated in redshift (as opposed to what we did for \redmagic\ galaxies), as we did not implement the redshift dependence of the magnification parameters for the weak lensing sample (Section~\ref{sect:fsmethod}}).

\subsubsection{Photo-$z$ estimates: SOMPZ}
The SOMPZ method uses spectroscopic and multi-band photometric information, and data from a number of deep fields \citep*{y3-deepfields} where additional photometry in the infrared bands $YJKs$ and $u$-band is available, besides the the standard 5-band (\textit{grizY}) photometry available in the DES wide field. This additional information is used to break the degeneracies in the photo-$z$ estimates of the DES wide field galaxies (which have fewer bands available). This is achieved by creating {two} Self-Organizing Maps \citep[SOM,][]{kohonensom}, {one mapping the deep/spectroscopic galaxies into a 2D grid of cells using their 8-band fluxes, and another mapping the WL sample galaxies into a 2D grid using the $riz$ photometry. A probabilistic mapping from the wide-field SOM to the deep-field SOM is generated using the ``Balrog'' source-injection simulations} \citep{y3-balrog} {and a map from the deep-field SOM to redshift is estimated using the spectroscopic data.}
\gbso{for the spectroscopic, deep and wide fields; a mapping of the spectroscopic redshifts to the deep SOM is first obtained and then it is followed by a mapping from the deep to the wide SOM.}

The tomographic bins are constructed as follows: a first set of edge values are arbitrarily selected. Each galaxy of the redshift sample is then assigned to the tomographic bin in which its redshift estimate falls. A number of galaxies at this point share the same photometry cell of the wide-field SOM and same tomographic bin, so the cell in its entirety is assigned to the bin to which the majority of its galaxies live. The initial bin edges are adjusted to yield approximately the same number of galaxies, and finally the whole procedure is repeated with the new bin edges. After completing this procedure, the final bin edges are [0.0, 0.358, 0.631, 0.872, 2.0] for the Y3 weak lensing source catalogue.

The full Y3 SOMPZ procedure is described in \cite*{y3-sompz}. A number of \gbso{different} factors contribute to the \gbso{total} error budget of the method: 1) shot noise (i.e., the limited number of galaxy redshifts available); 2) sample variance (i.e., the fact that the spectroscopic and deep fields span a limited area); 3) systematic uncertainties in the spectroscopic/multi-band photometry samples; 4) uncertainty in the methodology in general; 5) photometric calibration uncertainties in the Y3 deep fields, i.e. the uncertainty on the zero-point calibration in each band.

The total error budget is dominated by the photometric calibration uncertainty in the low redshift bin, while it is dominated by sample variance and biases in the spectroscopic/multi-band photometric samples in the high redshift bins \citep*{y3-sompz}.

The SOMPZ method {incorporates methods for assessing the likelihood $\mathcal{L}[ \textrm{PZ} | n_u(z)]$ of obtaining the various SOMPZ data elements (SOM cell counts, etc.) given a candidate set of $n_u(z)$ redshift distributions for the tomographic bins, which account for shot noise and sample variance in the various catalogs used by SOMPZ.  The construction of this likelihood, and the methods for sampling candidate $n(z)$ distributions from it, are given by} \citet{Sanchez2019}.
\gbso{provides a number of N(z) realisations encompassing the uncertainties related to shot noise, sample variance and uncertainties in the methodology. As for the data, the method also takes into account uncertainties related to biases in the spectroscopic/multi-band photometric samples} {Potential selection biases in the spectroscopic redshift assignments are estimated} by compiling $n(z)$ realisations obtained by {calibrating with} three different sets of spectroscopic/multi-band photometric samples. Redshift uncertainties related to the zeropoint calibration are added \textit{after} the SOMPZ realisations are informed by the clustering measurements \citep*{y3-sompz}. This is done for efficiency reasons and it does not affect the main results of this work.

The SOMPZ process is {completely reproduced} in simulations, {including the creation of spectroscopic catalogs from small-area surveys, but these simulations} do not take into account the uncertainties related to {unknown redshift selection biases}
in the spectroscopic/multi-band samples. As result of the slight differences of the simulated Y3 source sample data equivalent, the bin edges in the equivalent Buzzard catalogue are [0.0, 0.346, 0.628, 0.832, 2.0]. Estimates of the $n(z)$ obtained in simulations are shown in Fig.~\ref{fig:Nz_sims_data}. 

\section{Results on simulations and systematic errors}\label{sect:sims_tests}
\begin{figure*}
\begin{center}
\includegraphics[width=0.95 \textwidth]{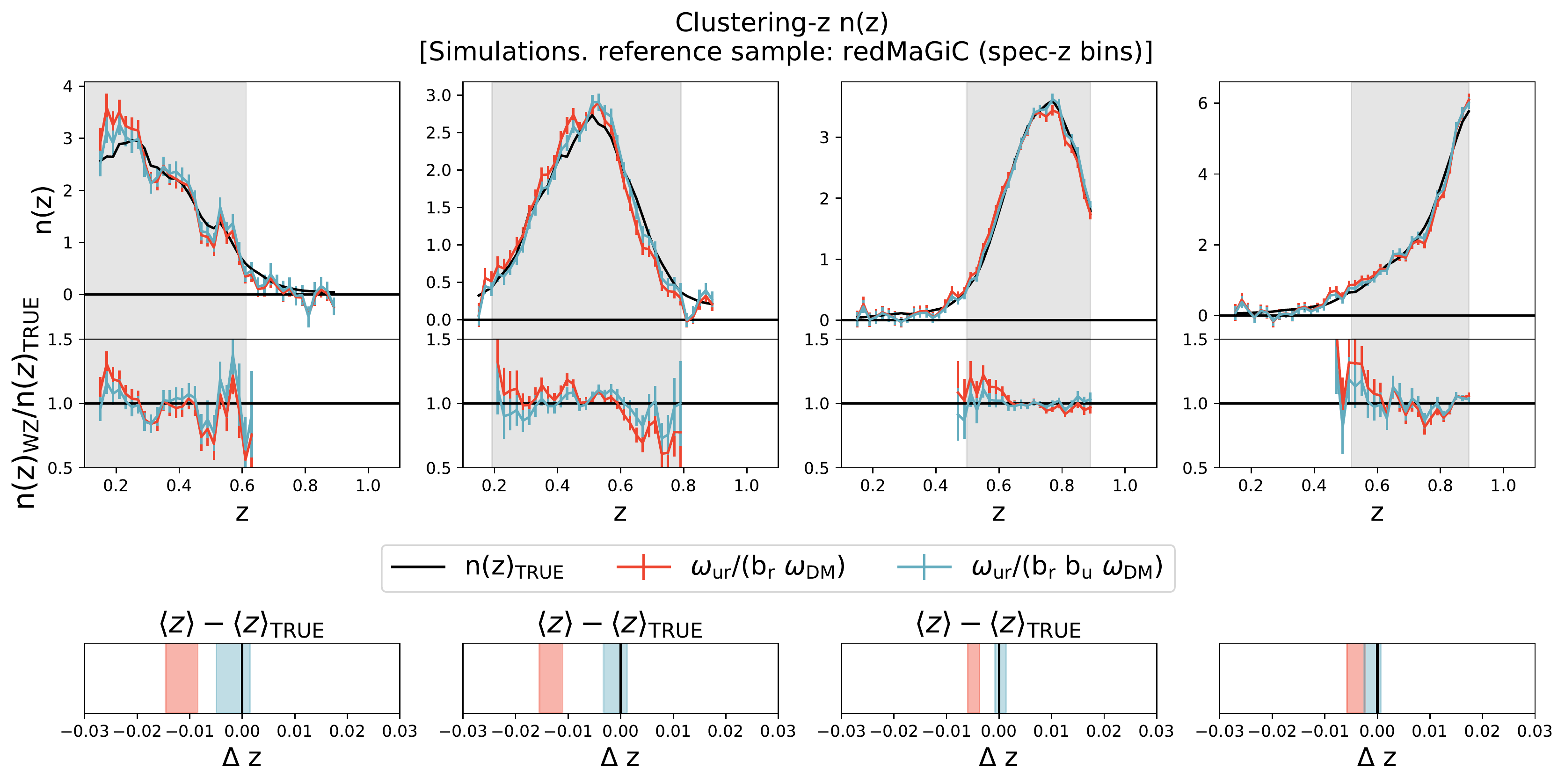}
\end{center}
\caption{Sources redshift distributions estimated using clustering redshift in simulations \giulia{for an \textbf{idealised setup} (see text \S \ref{sect:sims_tests})}, compared to the true (black lines). Top panels show the redshift distributions; middle panels show the ratio between the true $n(z)$ and the $n(z)$ estimated using clustering redshift; bottom panels show the mean of the redshift distributions. Red lines represent the clustering redshift estimates obtained using the estimator introduced by Eq.~\ref{eq:final_estimator}. Blue lines represent the clustering redshift estimated obtained further correcting for the term $\bu$, which is only possible in simulations. The four different tomographic bins used in the DES Y3 cosmological analysis are shown. We used \redmagic\ galaxies as the reference sample, binned using true redshifts. We also subtracted from the clustering redshift $n(z)$ estimates the expected magnification contribution in simulations. The redshift distributions are normalised over the same interval. Grey shaded regions indicate the interval considered for the mean matching method. The mean of the distributions showed in the bottom panels are computed only considering the grey intervals. \marco{Error bars only include statistical uncertainties}.}
\label{fig:nz_true}
\end{figure*}


In this section we present the results of our two calibration strategies performed in simulations. In particular, we aim to evaluate the systematic uncertainties of each method, and verify that the calibration procedure in simulations works as expected. {Note that at no point are the simulations used to make corrections to the data; rather the simulations are used to (1) estimate the level of uncertainty to assign to various systematic errors, and (2) validate that the method yelds results for $n(z)$ consistent with truth.}

Before focusing on the details of the two calibration procedures, we show in Fig.~\ref{fig:nz_true} the redshift distributions estimates obtained using the clustering-based $n_u(z)$ estimator (following Eqs.~\ref{system}, \ref{eq:final_estimator}) on simulations, compared to the true distributions. The angular scales considered in the clustering measurements have been chosen to span the physical interval between 1.5 and 5.0~Mpc. {These bounds (which are applied to the data as well) are selected so that the upper bound is below the range used for the $w(\theta)$ statistics used in cosmological analyses, thus allowing the WZ likelihoods to be essentially statistically independent of cosmology, and permitting us to produce $n(z)$ samples in an MCMC chain that runs before, and independent of, the cosmology.
The values of $b_{\rm r}$ in the WZ analysis are not required to \giulia{match} those used in the cosmological analyses.
The lower bound is chosen to produce high signal-to-noise ratio $S/N$ while \marco{mitigating} failures of the linear bias model.}


We start with an idealised case: the distributions shown in Fig.~\ref{fig:nz_true} are obtained using \redmagic\ galaxies as a reference binned using true redshift. We have also subtracted magnification contributions to the clustering estimator, \marco{using the true $n(z)$ for the modelling} and the best-estimate values of the magnification coefficients \marco{$\alpha$ (Eq. \ref{eq:magnification})} for the \redmagic\ and the WL samples, as reported in \S \ref{sect:data}.

In simulations we also have an accurate estimate of $\bu(z)$, obtained from the auto-correlations of each of the tomographic bins of the unknown sample, divided into thin bins of width $\Delta z =0.02$\footnote{\marco{In order to measure the auto-correlations, we generated randoms properly accounting for the WL mask. We also created systematic weights for the WL sample using the same procedure used for \redmagic\ galaxies (although we found they have a negligible impact).}}. This is not possible in data since the precision of the photometric redshift is not sufficient to divide the sample in bins of adequate width. Fig.~\ref{fig:nz_true} shows the impact on the estimated \Nz's of assuming we know $\bu(z)$ with good accuracy (in cyan), dividing Eq.~\ref{eq:final_estimator} by $\bu(z)$. We note that correcting for $\bu$ drives both the shape of the distributions and the mean value closer to the truth, \marco{which are otherwise biased}. As we cannot estimate $\bu$ in data, this highlights that variation in $\bu$ introduces a systematic uncertainty that has to be quantified. \marco{Note that the errors bars in Fig.~\ref{fig:nz_true} \textit{only include statistical uncertainties}}.


In the following subsections, we tested the accuracy of our calibration procedure uing the two different approaches outlined in Section~\ref{sect:methodology}, i.e. the mean-matching and the full-shape methods.

\subsection{Method 1 (mean-matching): systematic uncertainties estimation in simulations}\label{sect:systematic_method1}

We test in this section the mean-matching clustering-based photo-$z$ calibration method. The metric used here to assess the accuracy of our methodology is the difference between the mean of the recovered redshift distribution and the true mean, as follows:
\begin{equation}
\Delta \avg{z} \equiv \vert \avg{z}_{\rm true}-\avg{z}_{\rm WZ} \vert.
\end{equation} 
%
{As described in \S\ref{sec:combination},
$\avg{z}$ is calculated over a restricted redshift interval} \gbso{of 2 standard deviations around the mean of the SOMPZ distributions, i.e.} $\avg{z}_{\rm SOMPZ} - 2\sigma_{\rm SOMPZ} < z < \avg{z}_{\rm SOMPZ} + 2\sigma_{\rm SOMPZ}$ {to reduce the impact of magnification.  The redshift intervals are of course also truncated at the bounds of the reference sample.  The same redshift intervals are used for simulations as for data (see Fig.~\ref{fig:nz_true}).} The \gbso{extrema of the} intervals used are $[0.14,0.62]$, $[0.18,0.80]$, $[0.46, 0.90]$, $[0.48, 0.90]$ for \redmagic\, and $[0.10,0.62]$, $[0.18,0.80]$, $[0.46, 0.98]$, $[0.48, 1.06]$ for BOSS/eBOSS. \gbso{Given the similarities between the SOMPZ distributions in simulations and in data (Fig.
~\ref{fig:Nz_sims_data}), we also used the same intervals when applying our methodology to the data.}

\subsubsection{Systematic uncertainties}
\begin{table*}

\caption {\textbf{Mean-matching method}, total systematic error budget for the two reference samples used in this work. We also report the contribution due to each single source of systematic uncertainty, as a function of tomographic bin. As for the \redmagic\ systematic, we also report in parentheses the values of the uncertainties we would have obtained if we had not included the correction factor in the bias estimation (see Section~\ref{sect:systematic_method1}. }
\centering
\begin{adjustbox}{width=0.9\textwidth}
\begin{tabular}{|c|c|c|c|c|}
\hline

\textbf{Systematic} & \textbf{tomo bin 1}& \textbf{tomo bin 2} & \textbf{tomo bin 3}& \textbf{tomo bin 4}\\
\hline
methodology:		& $0.002  \pm 0.003$    &$0.001 \pm 0.002$   &$0.000 \pm 0.001$    &$0.001 \pm 0.002$  \\  
magnification:		&0.004     &0.005   &0.003    &0.004  \\
WL galaxy bias unc:	&0.013             &0.013             &0.013             &0.013           \\
redMaGiC syst:	&0.000 (0.014)           &0.001 (0.007)          &0.002 (0.000)           &0.005 (0.003)           \\
\hline
\hline
\textbf{total systematic \redmagic}:		&0.014    &0.014    &0.014     &0.015  \\   
\textbf{statistical \redmagic}:		&0.003     &0.002    & 0.001    &0.002 \\  
\hline
total systematic BOSS/eBOSS:   &0.014    &0.014     &0.014     &0.014 \\

statistical BOSS/eBOSS:   &0.007    &0.006    & 0.004    & 0.006 \\

\hline
\end{tabular}
\end{adjustbox}
\label{table1}
\end{table*}
We quantify here the systematic uncertainties of the mean matching method. {Since the mean-matching method reduces each \Nz\ to its windowed mean $\avg{z},$ the systematic errors will be quantified by the uncertainties that they imply should be added (in quadrature) to the $\sigma_z$ values of Eq.~\ref{likeY1}.} We note that the absolute value of the terms in Eq.~\ref{eq:final_estimator} are irrelevant for this method, as we are only interested in how they evolve with redshift. 
In principle, in the absence of magnification, assuming perfect reference sample redshift accuracy (e.g., \redmagic\ redshifts to be exact), assuming that we are able to successfully estimate all the terms in Eq.~\ref{eq:final_estimator}, and assuming that we know the galaxy-matter bias evolution of the unknown sample, we should correctly recover the mean of the unknown redshift distributions. The above assumptions might not hold when applying this methodology in data, causing a systematic bias in the calibration, In particular, $\Delta \avg{z}$ can differ from zero because of the following reasons:
\begin{itemize}
    \item 1) the approximations that allowed us to factorize the integral in Eq.~\ref{crosscorr} into ${b}_{\rm r}(z)  {b}_{\rm u}(z){w}_{\rm DM}(z)$, might not hold (e.g., linear bias model, infinitesimally thin bins), leading to inaccuracies in the modelling at small scales. We will quote these as \textbf{methodology systematics}. \marco{This systematic does not depend on the reference sample used}.
    \item 2) Magnification contribution. In the mean matching approach, we do not correct for magnification effects, as we cut the tails of the redshift distributions. This systematic quantifies how effective our cut is. We will refer to this has \textbf{magnification systematic}. \marco{This systematic depends on the particular reference sample used, but in the following we will make an order-of-magnitude estimate that should be valid for both reference samples}.
    \item 3) The clustering-based estimator ignores the redshift evolution of the galaxy-matter bias of the unknown sample (\textbf{WL galaxy bias uncertainty}).  \marco{This systematic does not depend on the reference sample used}.
    \item 4) The reference sample is binned using photometric redshifts and not spectroscopic redshifts. This only applies to the \redmagic\ case. We will refer to this as \textbf{\redmagic\ systematic}.
    
\end{itemize} 
We studied the performance of the estimator described in Eq.~\ref{eq:final_estimator} for four cases, starting from an ideal environment free from the effects of systematics and introducing one uncertainty at the time, leading to a more complex, realistic case. This allows us to estimate separately the magnitude of each systematic independently. In the following tests, we will only use the \redmagic\ galaxies as a reference sample to estimate the systematic uncertainties. Indeed, the BOSS/eBOSS sample should be affected by the same systematic uncertainties as the \redmagic\ sample, except for the \redmagic\ systematic. 

We begin with the most ideal case possible, shown in Fig.~\ref{fig:nz_true}, which we already described at the beginning of this section. Recall that for this case we used \redmagic\ galaxies as a reference binned using true redshifts, we corrected for the bias evolution of the unknown and reference sample, we corrected for the redshift evolution of the clustering of dark matter, and we subtracted magnification effects\footnote{This is achieved by subtracting the expected magnification signal from the measured cross-correlations $w_{\rm ur}$.} assuming the fiducial values of $\alpha_{\rm r}$ and $\alpha_{\rm u}$. The $\Delta \avg{z}$ mean for this case provides an estimate of the methodology systematic, and it is reported in the first line of Table \ref{table1}. This value is compatible with zero within statistical uncertainty \marco{(estimated through jackknife resampling)}, indicating that for the scales considered in this work (1.5--5.0~Mpc), the approximation of linear bias model, and infinitesimally thin redshift bins are good enough for the purpose of calibrating the mean with clustering information.


We next estimate the systematic uncertainty due to magnification effects \marco{(we remind the reader that in the mean matching method we do not model magnification effects)}. When adding back the contribution due to magnification that we subtracted in the preceding case, we obtain an increase of $\Delta \avg{z} \approx 0.002$. To be more conservative, we \gbso{can try to} estimate the impact on $\Delta \avg{z}$ if the data had different (and potentially larger) values of $\alpha_{\rm r}$ and $\alpha_{\rm u}$ than the ones estimated in simulations. We compute the magnification term $M(\theta)$ assuming Gaussian priors $\alpha_{\rm r} \sim \mathcal{N}(0,2)$ and $\alpha_{\rm u} \sim \mathcal{N}(0,2)$, and measure the resultant scatter in $\Delta \avg{z}$. \marco{These priors are rather wide, but even with these broad priors, magnification is a negligible component of our final error model}. Indeed, we obtain an RMS scatter on this metric \gbso{of the order} of $\Delta \avg{z}_{\rm RMS} = (0.004,0.005,0.003,0.004)$ for the four tomographic bins. We note that these values are up to a factor 10 smaller than what we would have obtained by including the tails of the redshift distributions, justifying the cut we introduced at the beginning of this section. {These values, in the second row of Table~\ref{table1}, are taken as the magnification contribution to $\sigma_z.$}

  We next quantify the impact \gbso{on the mean redshift} of ignoring the redshift evolution of the galaxy-matter bias of the unknown sample $\bu(z)$, as this cannot be measured in data. \gbso{Building up from the precedent case, we estimate our redshift distributions without correcting for the bias evolution of the unknown sample, and look at the difference in the mean redshift compared to the previous case. The impact on $\Delta \avg{z}$ is quantified in Table \ref{table1}, and reads, for the four tomographic bins,} {We estimate the size of this effect in the simulation by assuming a constant $\bu$ for each tomographic bin, and we obtain the resultant shifts of} $\Delta \avg{z}$ of  $(0.010,0.013,0.006,0.001)$. The effects of redshift-dependent $\bu(z)$ on the mean and on the shape of the clustering-$z$ \Nz\ \gbso{can also be appreciated from} {are shown in}  Fig. \ref{fig:nz_true}: \gbso{as the two clustering redshift \Nz\ estimates only differ because of} {the red and blue values differ only in the presence of the $\bu(z)$ term in the latter.}  {Given that the} \gbso{Also in this case we decide to take a more conservative approach, moved by the fact that the} WL galaxy bias uncertainty is the dominant uncertainty of the WZ method,
{we take the conservative approach of assigning an RMS systematic value to every bin that is}
\gbso{We assume the WL galaxy bias uncertainty to be the same in each tomographic bin, and} equal to largest $\Delta\avg{z}$ found in Buzzard, i.e., $\Delta \avg{z}=0.013$ estimated for the second bin. 
{This $\sigma_z$ contribution is listed in the third row of Table~\ref{table1}.}

Finally, we estimate the systematic uncertainty {in $\avg{z}$} due to {inaccuracies in the bin-shape integral in Eq.~\ref{autocorrelation1} for \redmagic\ galaxies when they are placed}
\gbso{the fact that \redmagic\ galaxies are binned} into thin bins using their photo-$z$ estimates. This is done {in the simulation} by comparing the $\avg{z}$ estimates obtained when binning the \redmagic\ galaxies using true redshifts \gbso{(as in the previous case)} to estimates obtained when binning using \redmagic\ photo-$z$. The photo-$z$ accuracies of \redmagic\ galaxies is better than those of the weak lensing sample, but not as good as those of a spectroscopic sample. This can introduce {two kinds of errors in $\avg{z}$: first,} \gbso{some uncertainties in the recovered redshift distributions. E.g.,} if all \redmagic\ photo-$z$ estimates were biased towards lower redshift, we would {infer a similarly biased \Nz.} \gbso{expect the clustering redshift $N(z)$ to be similarly biased. Due to the non negligible scatter of \redmagic\ photo-$z$, we also expect the clustering-$z$ $N(z)$ to be smoother compared to the case where the true redshift is used to bin the reference sample, as we cannot capture fluctuations in the true $N(z)$ over scales smaller than the intrinsic \redmagic\ photo-$z$ uncertainty.} {Second, the change in shape or width of the $n_{\rm r}(z)$ because of photo-$z$ errors can cause $\int {\rm dz} n_{\rm r}(z)^2$ to be wrong which propagates to a shift in $\avg{z}$.}


The shifts $\Delta\avg{z}$ that result from binning the \redmagic\ galaxies using \redmagic\ photo-$z$ rather than true redshifts are given in the 4th row of Table~\ref{table1}. \gbso{The values of the \redmagic\ uncertainties are reported in Table \ref{table1}.} We do not report statistical uncertainties, as they are negligible, since the shifts are computed taking the difference of two highly correlated measurements. The shifts are relatively small {and unimportant in comparison to the $b_{\rm u}$ uncertainties.} We also report in parentheses the \gbso{values of the uncertainties}{errors} in $\avg{z}$ we would have obtained had we not included the correction factor  of Eq. \ref{autocorrelation1} when estimating the galaxy-matter bias of \redmagic\ galaxies. Given the difference between the two estimates, {the correction due to the $n_{\rm r}^2(z)$ integral clearly}  cannot be neglected when applying the methodology to data. \marco{Last, we also estimated the \redmagic\ $\Delta\avg{z}$ using theory data vectors of the cross-correlation signal $w_{\rm ur}$, and modelling the \redmagic\ redshift distributions in each reference bin assuming the \redmagic\ photo-$z$ uncertainties estimated from data (Fig.~\ref{fig:redmagic_unc}), rather than the ones from the Buzzard simulation. This test  delivered $\Delta\avg{z}$ of the same order of magnitude as the ones estimated directly in Buzzard and reported in Table~\ref{table1}.}

Before reporting the total error budget for the mean matching method, we validate the assumption that we can assume a
fixed cosmology when calculating the clustering of dark matter, $w_{\rm DM} (z)$.  Assuming different values for the cosmological parameters ($\Omega_{\rm m} = 0.4$, $\sigma_8=0.7$) results in a negligible shift, $\Delta \avg{z} < 10^{-3}$.

\gbso{The systematic uncertainties evaluation done in this section is key to understand which are the systematics clustering methods are most sensitive to.} The total error budget is reported at the end of Table  \ref{table1}, and \gbso{it} is obtained \gbso{following the procedure adopted in DES Y1 analysis} 
  by adding in quadrature all the single sources of errors, assuming they are independent. \gbso{As we already mentioned} The dominant source of uncertainty is \gbso{due to} the potential redshift evolution of the weak lensing sample, which we do not model in \gbso{the current analysis} {the mean-matching analysis of the real data nor in the validation analyses of the simulations, which are described next.}





\subsubsection{Application of the method in simulations}

\begin{figure*}
\begin{center}
\includegraphics[width=0.95 \textwidth]{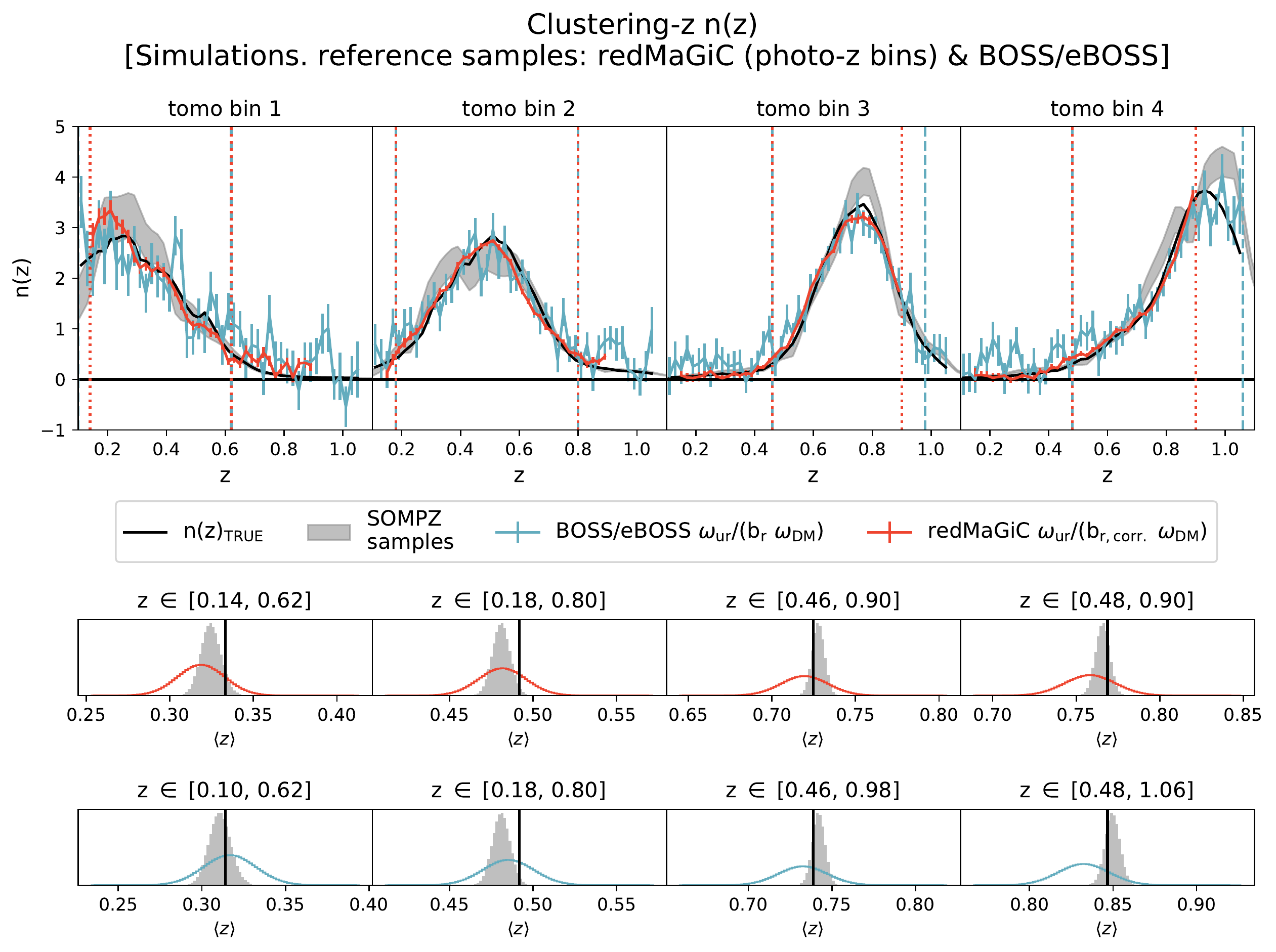}
\end{center}
\caption{\textit{Upper panels}: the redshift distributions estimated per Eq.~\ref{eq:final_estimator} {for a {realistic setup} (see text \S \ref{sect:sims_tests})}, compared to the truth (black lines). We show both the redshift distributions obtained using \redmagic\ galaxies as a reference sample, binned using their \redmagic\ photo-$z$ estimates, and the ones obtained using BOSS/eBOSS galaxies as a reference sample. Magnification effects have not been subtracted here.  The grey bands show, as a comparison, the 1 $\sigma$ region encompassed by the SOMPZ realisations. Vertical dotted (dashed) lines indicate the intervals where the windowed means of the \redmagic\ (BOSS/eBOSS) have been computed. \textit{Central and bottom panels}: windowed mean of the redshift distributions. \marco{The WZ estimates are represented by Gaussian histograms with mean equal to $\langle z \rangle_{\rm WZ}$ and $\sigma$ equal to the uncertainty of the method. The error budget of the WZ mean redshift estimates includes both statistical and systematic uncertainties (estimated in \S \ref{sect:systematic_method1} and reported in Table \ref{table1}), contrary to what was shown in Fig. \ref{fig:nz_true} that only reported statistical uncertainties.} } 
\label{fig:nz_rmg_eboss}
\end{figure*}

\begin{figure}
\begin{center}
\includegraphics[width=0.45 \textwidth]{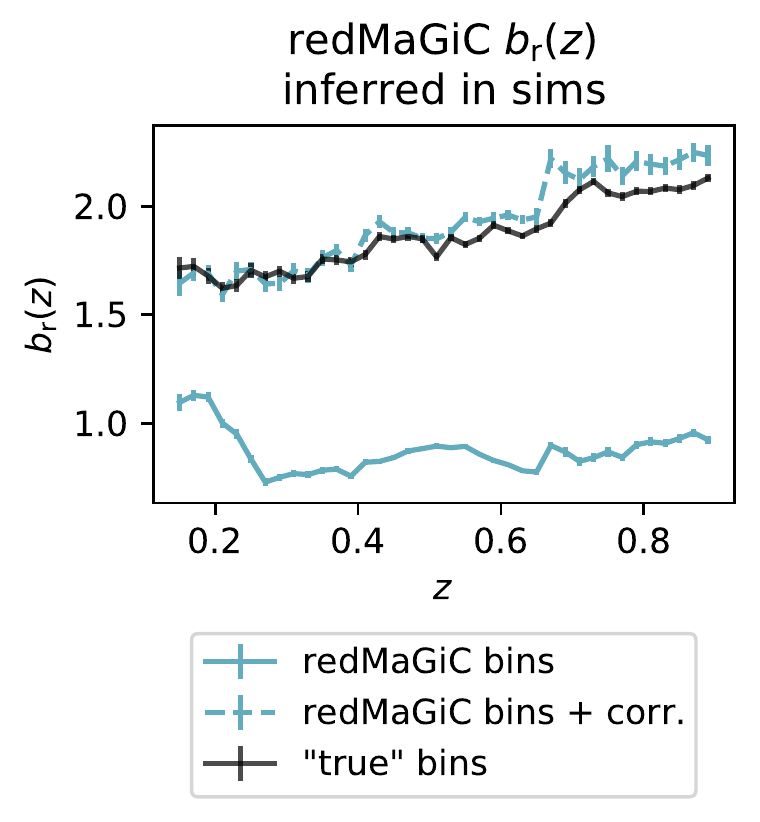}
\end{center}
\caption{Redshift evolution of the galaxy-matter bias $b_{\rm r}$ of simulated \redmagic\ galaxies, estimated with different binning. In particular, the black line has been obtained binning \redmagic\ galaxies using the true redshift, the solid light blue line  has been obtained binning \redmagic\ galaxies using \redmagic\ photo-$z$. The lower amplitude is due to the larger effective bin width due to the photometric uncertainties. The light blue dashed lines is computed from the light blue solid line after correcting for the larger width of the bins, following Eq. \ref{autocorrelation1}.}
\label{fig:bias_correction_sims}
\end{figure}

\marco{In order to apply the mean matching method in simulations, we run our clustering measurements using a realistic setup, for the two reference samples considered in this work. Fig.~\ref{fig:nz_rmg_eboss} compares the \Nz\ distributions obtained {from simulations} with \redmagic\ and  BOSS/eBOSS as reference samples. In particular,  \redmagic\ galaxies have been binned  using the \redmagic\ photo-$z$ estimates rather than the true redshifts and we did not correct for the bias evolution of the unknown sample. \marco{Magnification effects have not been modelled nor subtracted (although they are expected to have a small impact, due to our cut on the tails, as shown in the previous subsection)}. This plot highlights the differences between the two samples: \redmagic\ has a smaller statistical uncertainty, but the BOSS/eBOSS sample has a wider coverage in redshift, helping especially at higher $z.$ The distributions are compatible within errors. We note that in order to correct for the bias evolution of the reference sample when using \redmagic\ galaxies as a reference, we have to apply a correction to the width of \redmagic\ bins, as described in Eq. \ref{autocorrelation1}, to account for the broader distributions that \redmagic\ bins have compared to a top-hat bin. This correction is shown in Fig.~\ref{fig:bias_correction_sims}.}

\label{sec:meanmatch_validation}
\begin{figure*}
\begin{center}
\includegraphics[width=0.95 \textwidth]{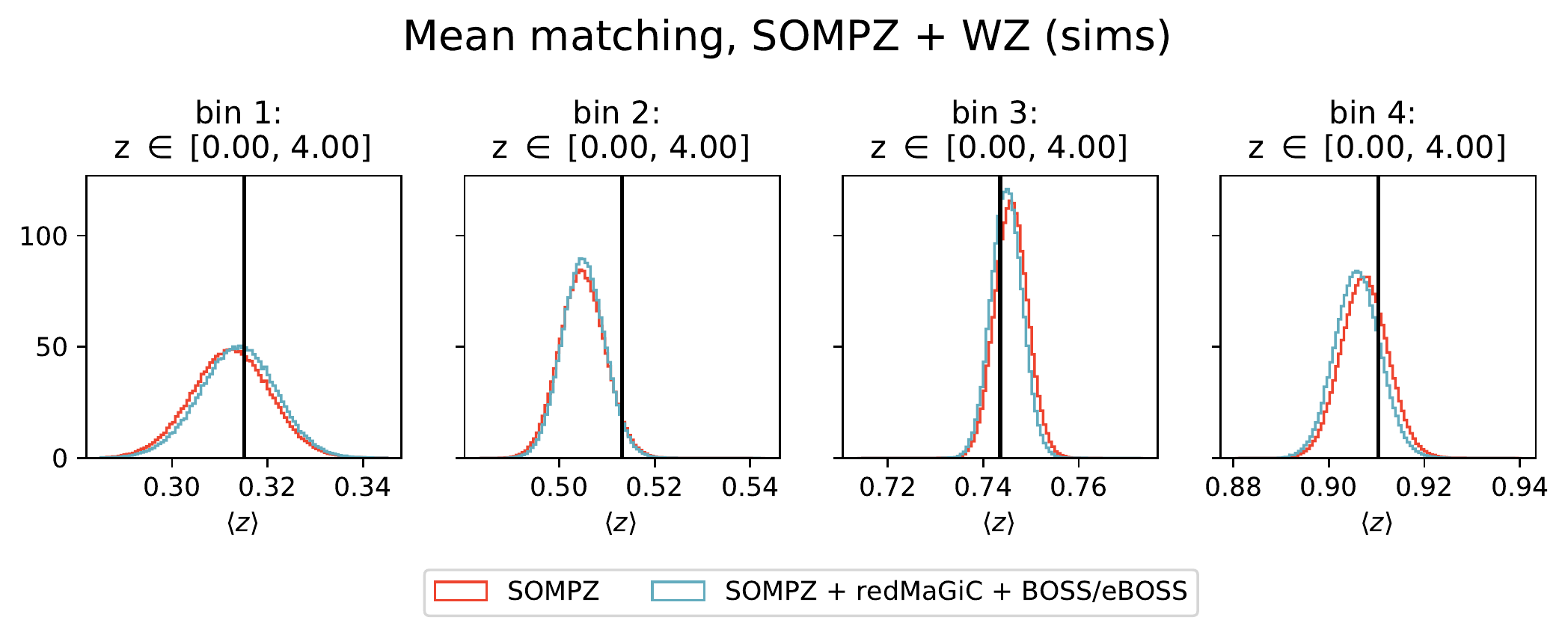}
\end{center}
\caption{Mean redshift posteriors for the 4 tomographic bins obtained using the mean matching method in simulations. Red histograms represent the distribution of the mean redshift of the SOMPZ realisations, whereas light-blue histograms show the mean redshift posteriors of the SOMPZ realisations using the WZ likelihood. The mean redshift of the SOMPZ realisations has been computed over a wide redshift interval ($0<z<4$), also including the redshift range where there is no WZ information.} 
\label{fig:dz_meanmatching_SOMPZ_WZ}
\end{figure*}

\marco{Once we have $n(z)$ WZ estimates, we first verify that the WZ windowed mean redshift estimates obtained using the two reference samples are both compatible within uncertainties (including systematic and statistical) with the truth, and with SOMPZ estimates. This is shown in the lower panel of  Fig.~\ref{fig:nz_rmg_eboss}. Note that the WZ windowed means are compatible by construction with the truth, given our modelling of the systematic uncertainties of the method.}

We can then proceed combining the WZ information with the SOMPZ method.  Recall that the SOMPZ method can provide samples of the $n_{\rm u}(z)$'s from its posterior distribution.  We can importance-sample these SOMPZ samples by assigning each a weight through the likelihood given by Eq.~\ref{likeY1}. As we have two reference samples, we \gbso{combined} multiply the likelihoods obtained using the \redmagic\ and BOSS/eBOSS samples; we assume the two likelihoods \gbso{were not completely independent, but that they} share the WL galaxy bias uncertainty but are otherwise \giulia{considered independent, which is a reasonable assumption given the fact the total error budget of the methodology is systematic dominated and the overlap between the two sample is minimal}.

{Fig.~\ref{fig:dz_meanmatching_SOMPZ_WZ}  shows, in red, the distributions of $\avg{z}$ over SOMPZ realisations, one panel for each tomographic bin.  Note that in this case, $\avg{z}$ is taken over $0<z<4,$ not restricted to narrower ranges where the WZ signal is measured and large.  The blue curves show the distributions of $\avg{z}$} \gbso{before and} after having being weighted by the WZ likelihood. \gbso{are shown in, compared to the true mean redshift.} The means and standard deviations of $\avg{z}$ of the SOMPZ realisations are also reported in Table~\ref{tableres}, with and without the importance weighting by mean-matching. {The importance-weighted $\avg{z}$ values are fully consistent with unweighted SOMPZ realizations, and with the truth for the simulations.}

\begin{table*}

\caption {\textbf{Simulations}. The mean redshift estimates of the SOMPZ distributions with and without clustering-based information, in simulations.}
\centering
\begin{adjustbox}{width=0.9\textwidth}
\begin{tabular}{|c|c|c|c|c|}
\hline

\textbf{case} & \textbf{tomo bin 1}& \textbf{tomo bin 2} & \textbf{tomo bin 3}& \textbf{tomo bin 4}\\
\hline
True $\langle z \rangle$:	&0.315 & 0.513  &0.743 & 0.910 \\ 
SOMPZ $\langle z \rangle$:	& $0.312 \pm 0.008$ & $0.505 \pm 0.005$ & $0.746 \pm 0.003$ & $0.907 \pm 0.005$ \\
 SOMPZ + WZ (mean-matching) :	& $0.314 \pm 0.008$ & $0.505 \pm 0.004$ & $0.745 \pm 0.003$ & $ 0.906 \pm 0.005 $ \\
SOMPZ + WZ (full-shape) :	& $0.312 \pm 0.009$ & $0.507 \pm 0.005$ & $0.747 \pm 0.004$ & $ 0.907 \pm 0.005 $ \\

\hline
\end{tabular}
\end{adjustbox}
\label{tableres}
\end{table*}

\marco{The WZ information in fact offers little improvement in the constraints from the SOMPZ realisation. The systematic errors we derive on $\avg{z}$ are larger than the statistical errors with DES Y3 data (Table~\ref{table1}), and also larger than the total errors estimated for the SOMPZ method (Fig.~\ref{fig:nz_rmg_eboss})}. This means that for the DES Y3 analysis, the mean-matching method can {be useful as an} independent cross-check of the SOMPZ methodology, but it does not significantly improve the constraints on the mean of the redshift distributions.

{This is not entirely surprising, because we have seen that the dominant systematic error in the mean-matching method (indeed for WZ in general) is the uncertainty in the redshift evolution of the bias of the unknown sample, $b_{\rm u}(z).$  Even a simple linear slope to $b_{\rm u}(z)$ will be imprinted on the inferred $n_{\rm u}(z)$ and shift $\avg{z}$, meaning that the dominant systematic error has its largest effect on this lowest-order moment of $n_{\rm u}(z)$.  Thus in some sense, $\avg{z}$ is the statistic for which we should expect WZ techniques to be least informative.  On the other hand, we expect $b_{\rm u}(z),$ and other sources of systematic error in the WZ method, to be smooth, low-order functions of $z.$ We will therefore look next into the ability of WZ data to constrain the full shape of $n_{\rm u}(z)$.}

%

\subsection{Method 2 (full-shape): systematic uncertainty estimation in simulations}\label{sect:systematic_method2}
\begin{figure*}
\begin{center}
\includegraphics[width=0.85 \textwidth]{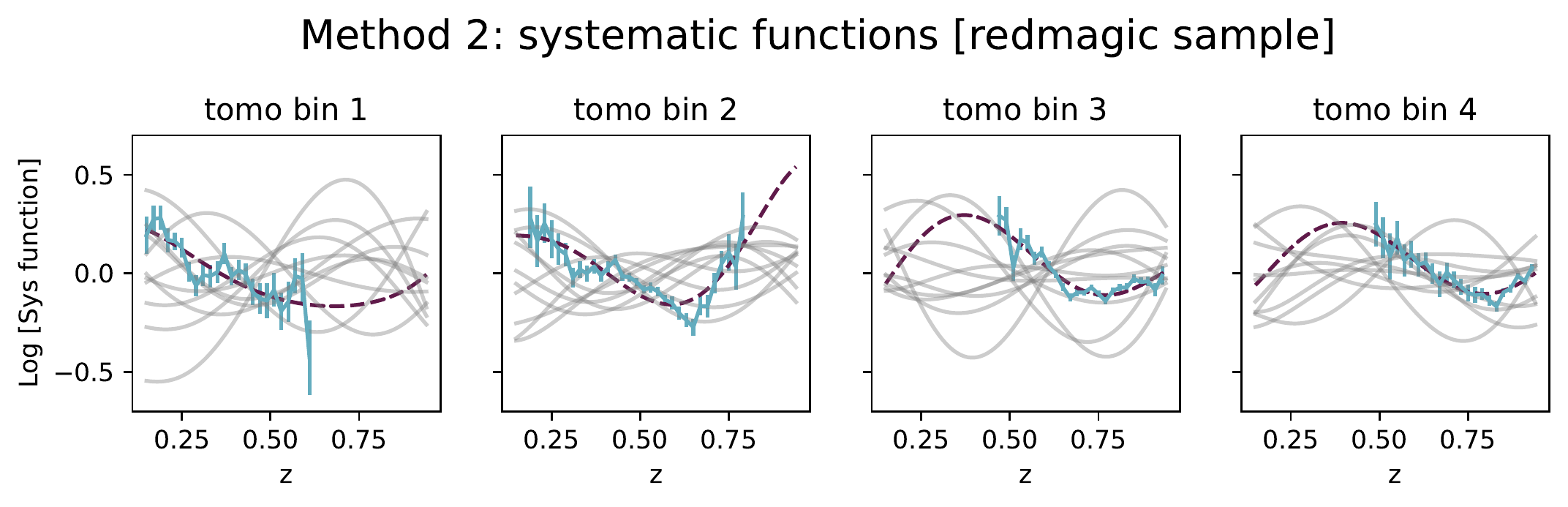}
\includegraphics[width=0.85 \textwidth]{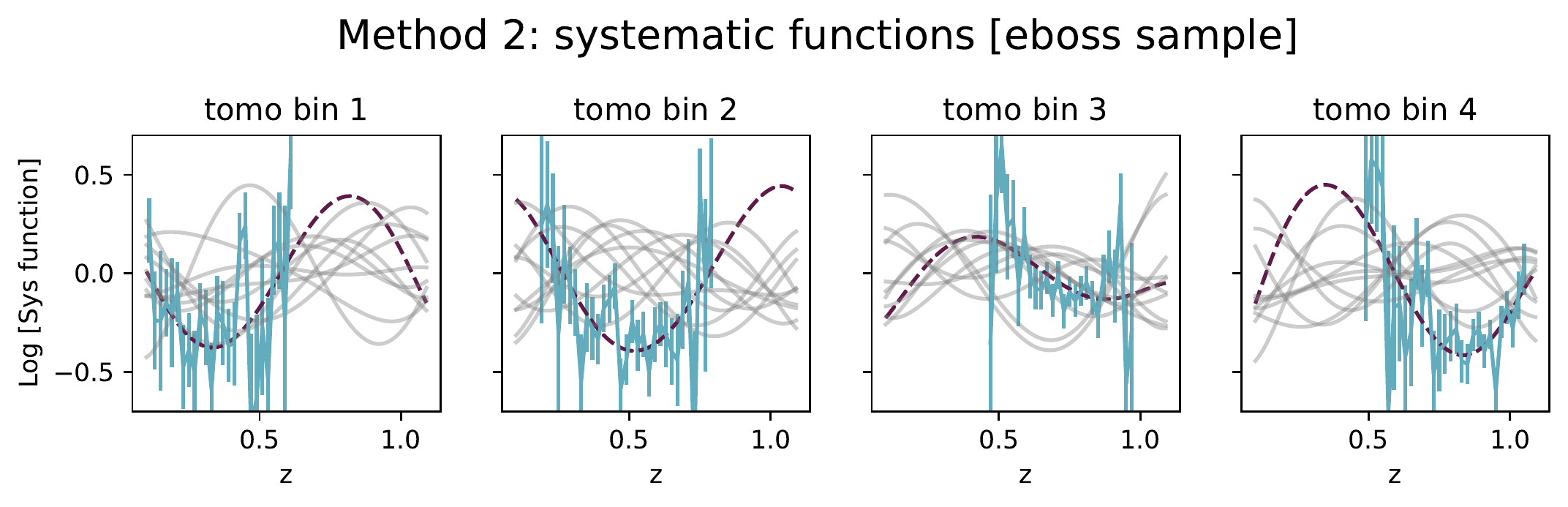}
\end{center}
\caption{Systematic uncertainties of the full-shape method as measured in simulations following Eq.~\ref{eq:def_syst_unc}, for the 4 tomographic bins and for the two reference samples considered (\redmagic\, upper panels, BOSS/eBOSS, lower panels). The measured systematic uncertainties are represented by the light blue lines; the purple dashed lines represent the best fitting model. The grey lines represent 10 random realisations of the systematic uncertainty model assumed for the full-shape method and described by Eq.~\ref{eq:model_syst_unc}.} 
\label{fig:systematic_method2}
\end{figure*}
{In the full-shape likelihood of \S\ref{sec:combination}, we produce a model for the $w(z)$ signal}
\gbso{We discuss here the second method investigated in this paper to combine the clustering information with the SOMPZ $N(z)$ realisations. This method involves forward modelling the full clustering signal} across the full redshift range covered by the reference samples (i.e., including the tails of the distributions) {and produce a likelihood for the observed $w(z)$ data. In practice, this allows us to constrain the full shape of the redshift distributions, not only the mean.  Here we use the Buzzard simulations to set the priors for the systematic-error parameters within this model.}


%

\subsubsection{Systematic uncertainty determination}

Recall that in \S\ref{sect:methodology}, specifically Eq.~\ref{wursys}, the cross-correlation signal is modelled starting from a proposed value for $n_{\rm u}(z)$ (e.g. provided by SOMPZ), the (measurable) reference-population properties $b_{\rm r}(z)$ and $\alpha_{\rm r}(z),$ and nuisance parameters for the (poorly known) bias and magnification properties of the source population $b_{\rm u}(z)$ and $\alpha_{\rm u}(z)$.
 We will set these last two as constant over redshift and marginalize over broad priors on these constants, to flexibly model the magnification signal.
The underlying function  ${w}_{\rm DM} (z)$ is estimated assuming a cosmological model. 

The final component of the $\hat w_{\rm ur}$ model is a function $\sys(z,\vecs)$ that multiplies the true clustering signal and will absorb the systematic errors described for the mean-matching method: failures of the linear-bias model itself; the unknown and redshift-dependent $b_{\rm u}(z)$; and possible errors in the $n_{\rm r}(z)$ functions for \redmagic\ bins.  The parameters \vecs\ of this systematic function will be marginalized as well, as per Eq.~\ref{eq:wzlike}.

Our strategy will be to determine what the $\sys(z)$ function is in the Buzzard simulation, and then produce a prior on the $\vecs$ parameters which allows marginalization over a broad family of functions with similar form of deviation from unity.  The $\sys(z)$ function is given substantial freedom for low-order, smooth variation with $z$, as we expect from all of the systematic errors, leaving the finer-scale information in $w_{\rm ur}(z)$ to constrain fine-scale behavior in $n_{\rm u}(z)$, i.e. the shape of $n_{\rm u}(z)$.

The blue data points in Fig.~\ref{fig:systematic_method2} plot the $\sys(z)$ functions observed in the Buzzard simulations, for both reference samples.  Namely they plot
\begin{equation}
\label{eq:def_syst_unc}
\sys_{\rm sim}(z_i) = \frac{{w}_{\rm ur}(z_i)-M(z_i)}{\hat{w}_{\rm ur}(z_i)-M(z_i)},
\end{equation}
where $M(z_i)$ is a nominal estimate of the magnification terms, and the model uses the true $n_{\rm u}(z), b_{\rm r}(z),$ and $n_{\rm r}(z)$ values.
We evaluate and plot this ratio only in the $z$ interval where the $w_{\rm ur}$ signal is large enough to have good signal-to-noise and subdominant magnification contribution. 
The \redmagic\ $w_{\rm ur}(z)$ uses \redmagic\ photo-$z$'s for binning, just as the real data do.

The $\sys_{\rm sim}$ ratio deviates from unity due to systematic effects, as expected.   We quantify this by the RMS 
of $\log\left[\sys_{\rm sim}(z)\right],$ which are measured to be (0.11, 0.07, 0.07, 0.11) for the \redmagic\ tomographic bins and  (0.18, 0.15, 0.10, 0.15) for BOSS/eBOSS.  From this we conservatively decide that the $\sys$ function needs to have the freedom to have RMS (log) fluctuations of $\approx0.15$ as $1\sigma$ deviations under its $p(\vecs)$ function.

We seek a parametric function $\sys(z;\vecs)$ and a prior $p(\vecs)$ which have these desired properties: 
\begin{itemize}
\item The function and prior yield a good fit to the $\sys_{\rm sim}$ measured in Buzzard.
\item The prior can be tuned to yield typical RMS variations in $\log\left[\sys(z)\right]$ at similar level to that seen in Buzzard.
\item The parametric form allows a similar smoothness of variation as seen in Buzzard, i.e. similar number of ``wiggles'' across the $0<z<1.2$ range where the WL source galaxies lie.
\item The RMS of $\log\left[\sys(z;\vecs)\right]$ as we vary \vecs\ under the prior $p(\vecs)$ is a flat function of $z$.
\item The prior on $\vecs$ is simple to construct and to use in a Hamiltonian Monte Carlo chain.
\end{itemize}

We chose the ${\rm Sys \left(z,\vecs\right)}$ function to be given by:
\begin{align}
\label{eq:model_syst_unc}
  log[\sys \left(z,\vecs\right)] & = \sum_{k=0}^M \frac{\sqrt{2k+1}}{0.85} s_k P_k(u), \\
  \label{eq:sys_rescale}
  u & \equiv 0.85 \frac{z-0.5(z_{\rm max}+z_{\rm min})}{(z_{\rm max}-z_{\rm min})/2}.
\end{align}
with $P_k(z_i)$ being the $k$-th Legendre polynomial, $M$ is the maximum order, and the second line linearly remaps the $z$ interval $[z_{\rm min},z_{\rm max}]$ to $[-0.85,0.85].$  The fraction under the summation makes the basis functions close to orthonormal so that the RMS of $\log(\sys)$ is $|\vecs|^2$.  The prior $p(\vecs)$ is chosen to be a simple diagonal normal distribution with standard deviations $\{\sigma_{s0},\ldots,\sigma_{sM}\}$ and means of zero.  
Mathematical details of this choice for the systematic function and its prior are given in Appendix~\ref{AppendixSys}.


A distinct set of nuisance parameters $\vecq=\{\vecp, \vecs\}$ (with $\vecq=\{b_{\rm u}^{'}, \alpha_{\rm u}^{'}\}$) are assigned to each combination of tomographic bin and reference sample, and each of these 8 sets of $w_{\rm ur}$ measurements are fit independently. 
We set $[z_{\rm min},z_{\rm max}]$ to span the full range of the reference catalog, $[0.14,0.90]$ for \redmagic\ and $[0.10,1.06]$ for BOSS/eBOSS.  We set $M=5$ and we set the $\sigma_{s_i}$ to yield an expectation value of 0.15 for the RMS of $\log\left[\sys(z)\right].$ \gary{The order $M$ was chosen by finding the value beyond which the RMS residual stopped decreasing for a fit of Eq.~\ref{eq:model_syst_unc} to the $\sys(z)$ function found in the simulated \redmagic\ $w_{ur}(z)$ data.  The
$\sigma_{s_i}$ prior is set to make the simulated $\sys(z)$ functions be $\approx 1\sigma$ fluctuations from a constant.}
Since $e^{s_0}$ is \gary{approximately} the mean bias of the unknown sample, \gary{and we expect the mean bias $b_r$ to be more uncertain than the variation with redshift,}  we treat the prior on $s_0$ somewhat differently, giving it a wide prior $\sigma_{s0}=0.6.$  The RMS of 0.15 is then allocated among the remaining elements $k\ge1$ of $\vecs$ which model redshift-dependent systematic errors.

The nuisance parameter $b_{\rm u}^{'}$ used in magnification estimation is given a Gaussian prior with $(\mu,\sigma)=(1.,1.5)$ (\marco{which encompasses the bias of the weak lensing sample as measured in simulation}).  The other magnification nuisance $\alpha_{\rm u}^{'}$ is given a mean estimated from image-injection simulations (\citealt{y3-balrog},Section~\ref{sect:data_wl}) and an uncertainty of $\sigma=1.$

The dashed curves in Fig.~\ref{fig:systematic_method2} plot the $\sys$ functions obtained from the maximum-posterior fits to the simulations' $w_{\rm ur}(z)$ data, combining the priors on the nuisance parameters with the likelihood of Eq.~\ref{eq:wzlike}.  All 8 cases are well fit, and the fitted functions remain well-behaved over the full $w_{\rm ur}$ redshift range even though the fit is done only for redshifts with strong signals.  We conclude that this formulation of the systematic errors is sufficient to model the systematic errors in our WZ measurement in the Buzzard simulation, and we assume that marginalization over $\vecq$ will allow us to capture the uncertainties present in the real data as well.

The grey curves in Fig.~\ref{fig:systematic_method2} show a few examples of $\sys(z;\vecs)$ functions obtained by random sampling of the prior $p(\vecs)$. This illustrates the flexibility of our model for the systematic uncertainty, which is able to model a large variety of curves. 

It is useful to ask whether this implementation of systematic errors in the full-shape method is consistent with the systematic uncertainties derived for the mean-matching method. This can be done by drawing many realisations of $\vecs$ from its prior, constructing a model $\hat w_{\rm ur}$ data vector using each realization of $\sys(z,\vecs),$ and then treating this model as data input to the mean-matching method.  Each realization of $\vecs$ then yields an estimate of $\Delta\avg{z}$ with respect to the true distribution. We obtained a typical $\left|\Delta \avg{z}\right|$ in the range 0.010--0.015 depending on the tomographic bin, in very good agreement with the total systematic uncertainties estimated in Table~\ref{table1} for the mean-matching method.




\subsubsection{Application of the method in simulations}\label{ss:ss}
\begin{figure*}
\begin{center}
\includegraphics[width=1 \textwidth]{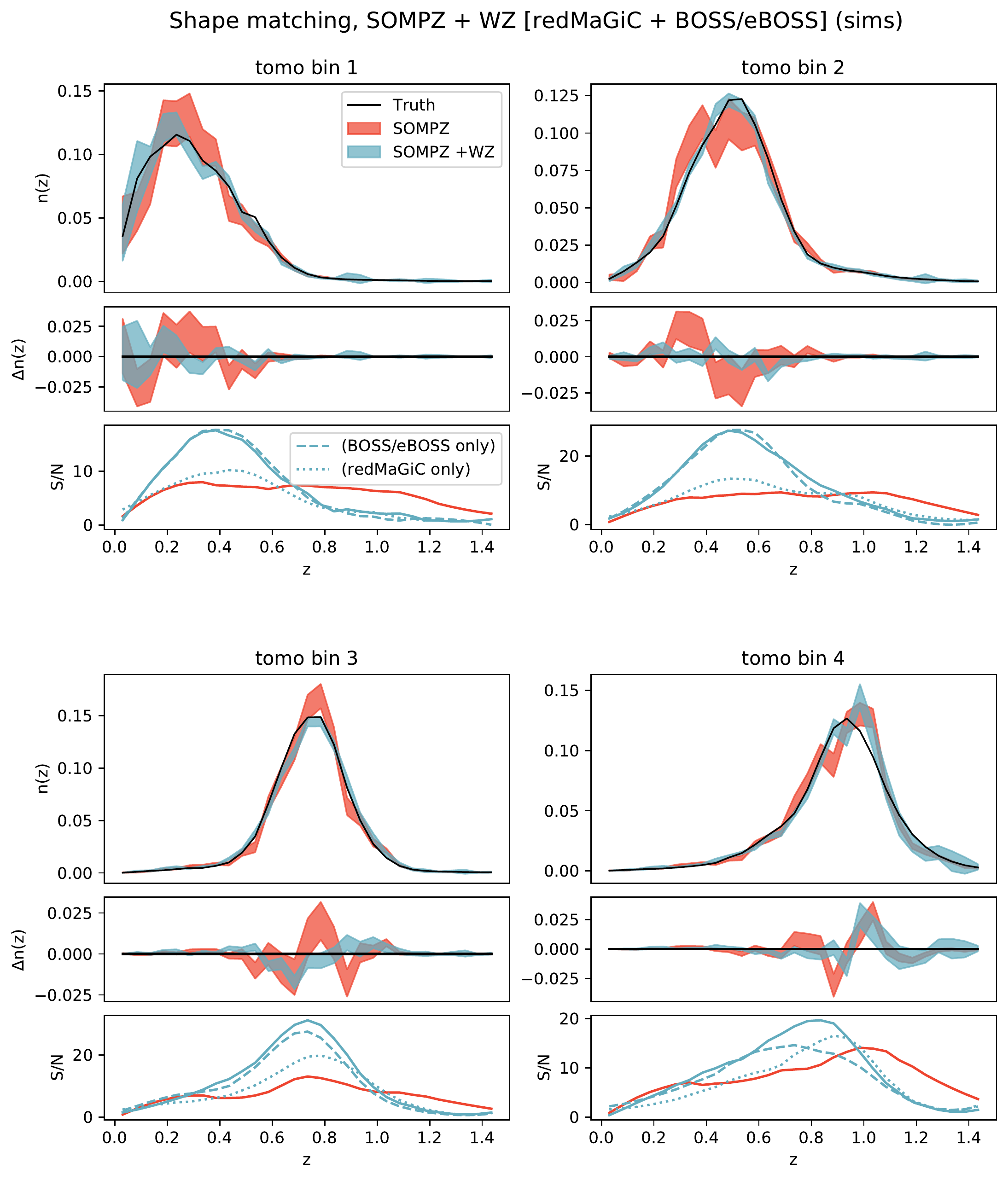}

\end{center}
\caption{For each tomographic bin, three panels are shown. \textit{Upper panels:} SOMPZ redshift distributions, as estimated in simulations, with and without clustering information (full-shape method). The bands encompass 68\% confidence interval of the SOMPZ $n(z)$ realisations. \textit{Central panels:} difference between the recovered $n(z)$ and the true $n(z)$ in simulations. \textit{Lower panels:} \marco{S/N, defined as the ratio between the SOMPZ  $n_{\rm u}(z)$ and its 68\% confidence interval of the SOMPZ realisations, with and without clustering information. The dashed (dotted) line has been obtained only using WZ constraints from \redmagic\ (BOSS/eBOSS) galaxies.}}
\label{fig:shape_match_sims_all}
\end{figure*}
Once our family of systematic functions is determined for the full-shape method, we may proceed to validating the performance of the combination of SOMPZ and the full-shape WZ method on the Buzzard simulations.  This combination is implemented (both in simulations and in data) by sampling the $n_{\rm u}(z)$ functions for all 4 tomographic bins from a posterior defined by the product of:
\begin{itemize}
\item the SOMPZ probability defined by \citet*{y3-sompz};
\item the WZ probability defined by Eq.~\ref{eq:wzlike} for the $w_{\rm ur}(z_i)$ measured against the \redmagic\ sample,  marginalized over $\vecq$ as described in Appendix~\ref{AppendixSys};
\item and likewise, the marginalized WZ probability derived for the BOSS/eBOSS sample, marginalized over $\vecq$ as described in Appendix~\ref{AppendixSys};
\end{itemize}
The WZ probabilities use $w_{\rm ur}(z)$ over the full redshift range of their respective reference samples. The reference-sample magnification coefficients $\alpha_{\rm r}$ and the cosmology used to derive $w_{\rm DM}(z)$ are held fixed to nominal values.  We verify below that the choices of $\alpha_{\rm r}$ and cosmology have insignificant impact on the outcome of the full-shape method. \marco{In this case, contrary to the mean-matching method, we consider the \redmagic\ and BOSS/eBOSS likelihoods independent, i.e. they do not share the WL galaxy bias uncertainty. We did this because in the full-shape case we did not split our systematic function into different source of errors, owing to an increasing complexity in the modelling. Given the flexibility of $\sys \left(z,\vecs\right)$ and the conservative choice on the RMS of $\log\left[\sys(z;\vecs)\right]$, considering the \redmagic\ and BOSS/eBOSS likelihoods independent should not be an issue for the methodology. } The sampling of the joint SOMPZ$+$WZ posterior is done using a Hamiltonian Monte Carlo method described in \citet{hmcpaper}.  


Fig.~\ref{fig:shape_match_sims_all} compares the Buzzard true redshift distribution to the distributions drawn from only the SOMPZ likelihood and the distributions drawn from the joint SOMPZ$+$WZ posterior.  The distributions of the mean redshifts per bin in the lower panels are not shown, but it is reported in Table~\ref{tableres}. It shows that the full-shape WZ likelihood adds little information on these mean $z$'s.  This is as we expect from the results and discussion of the mean-matching method in \S\ref{sec:meanmatch_validation}. The plots in \marco{Fig.~\ref{fig:shape_match_sims_all}, however, shows that the addition of full-shape WZ likelihood produces a remarkable improvement in the fidelity of the shape of $n_{\rm u}(z)$ to the truth. To better quantify the improvement, we also show the signal-to-noise (S/N) of the $n_{\rm u}(z)$ estimates, defined as the ratio between the SOMPZ  $n_{\rm u}(z)$ and the 68\% confidence interval of the SOMPZ realisations. The S/N is generally increased by the inclusion of the WZ information; in particular, the S/N is increased up to a factor of 3 in the relevant redshift range where $n(z)$ is substantially different from 0. In the same S/N panels of Fig.~\ref{fig:shape_match_sims_all}, we also show the contribution to the S/N increment due to \redmagic\ galaxies or BOSS/eBOSS galaxies alone. The latter sample mostly contributes in the redshift range $0.8<z<1.0$, whereas most of the WZ information at lower redshift comes from \redmagic\ galaxies.}

The SOMPZ method has strong fine-scale fluctuations in $n_{\rm u}(z)$ due to sample variance on the small regions of sky used for its deep imaging and spectroscopy. The WZ correlation functions, on the other hand, are measured over the full DES Y3 footprint and have high $S/N$ level. Although the clustering signal has a strong systematic uncertainty from the unknown WL bias, this systematic is slowly varying as a function of redshift and has less fine-scale fluctuations.  The WZ likelihood is thus able to drive the $n_{\rm u}(z)$ outputs to a smooth distribution, at least over redshifts where WZ reference samples are available.

\marco{We remind the reader that the clustering information alone cannot be used to infer the $n_{\rm u}(z)$, as the reference samples used in this work do not span the whole redshift range relevant for the DES Y3 $n_{\rm u}(z)$. Nonetheless, we can try to understand in simulations if the full-shape method would be unbiased independently of the SOMPZ information. We did this by importance-sampling realisations of the true $n_{\rm u}(z)$s shifted around their mean redshift, and by assigning to each sample a weight through the likelihood given by Eq.~\ref{eq:wzlike}. This test allowed us to recover the true $n_{\rm u}(z)$ within uncertainties, hence proving the method to be unbiased.}

Finally, we verify that the choices of the the parameters $\alpha_{\rm r}$ or the cosmology assumed to compute $w_{\rm DM}$ do not impact the methodology. We find that assuming different values for the cosmological parameters ( $\Omega_{\rm m} = 0.4$, $\sigma_8=0.7$) results in a shift in $\Delta \avg{z} < 10^{-3}$ on the calibrated SOMPZ redshift distributions. Concerning magnification, in order to roughly asses the impact of the exact values of the magnification coefficients $\alpha_{\rm r}$,  we verified that assuming values for $\alpha_{\rm r}$ that are $-1\times$ the fiducial ones resulted in shifts $\Delta \avg{z} < 10^{-3}$.   We conclude that the full-shape likelihoods, like the mean-matching, can be calculated in advance of and independent from the cosmology chains.

\section{Application to data}\label{sect:calibration}
\begin{figure*}
\begin{center}
 \includegraphics[width=0.95 \textwidth]{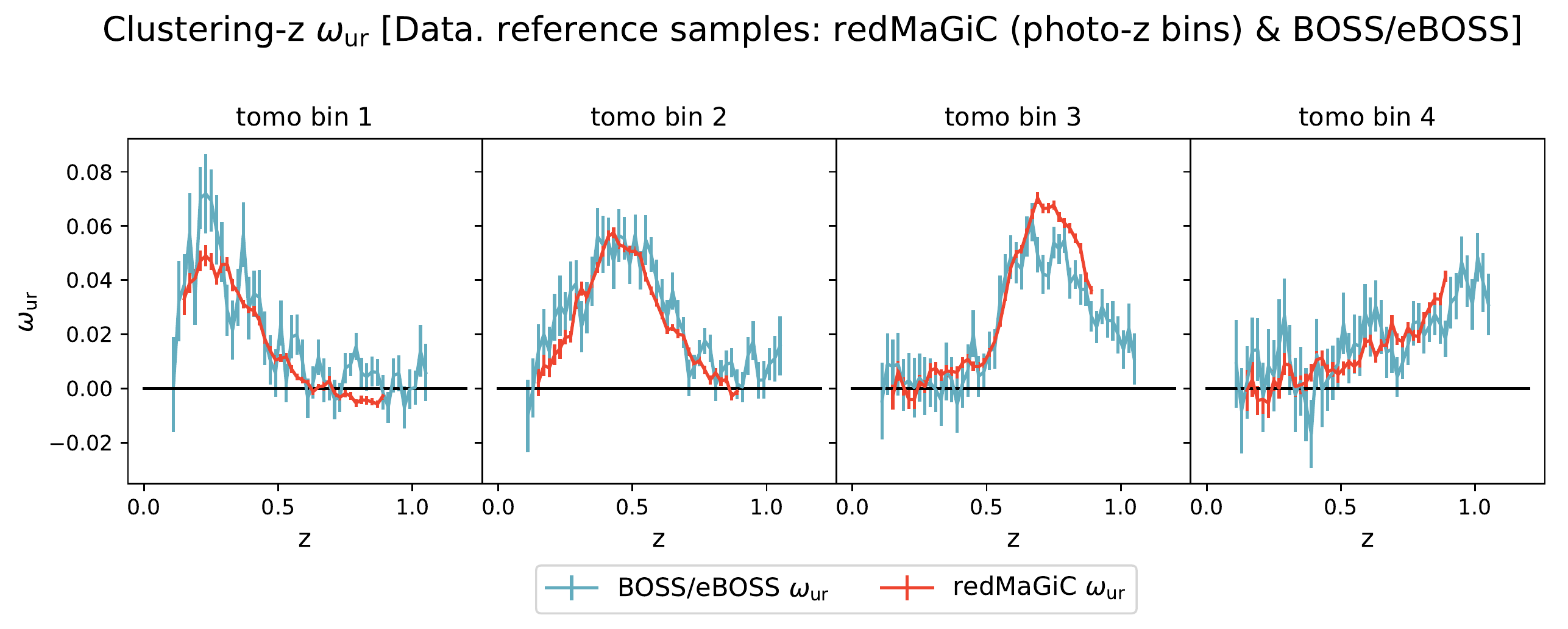}
\end{center}
\caption{The measured $w_{\rm ur}(z)$ for the DES Y3 data are plotted for each of the four tomographic bins, using reference samples from BOSS/eBOSS (blue) and \redmagic\ (red). }
\label{fig:w_data}
\end{figure*}

\begin{figure*}
\begin{center}
\includegraphics[width=0.95 \textwidth]{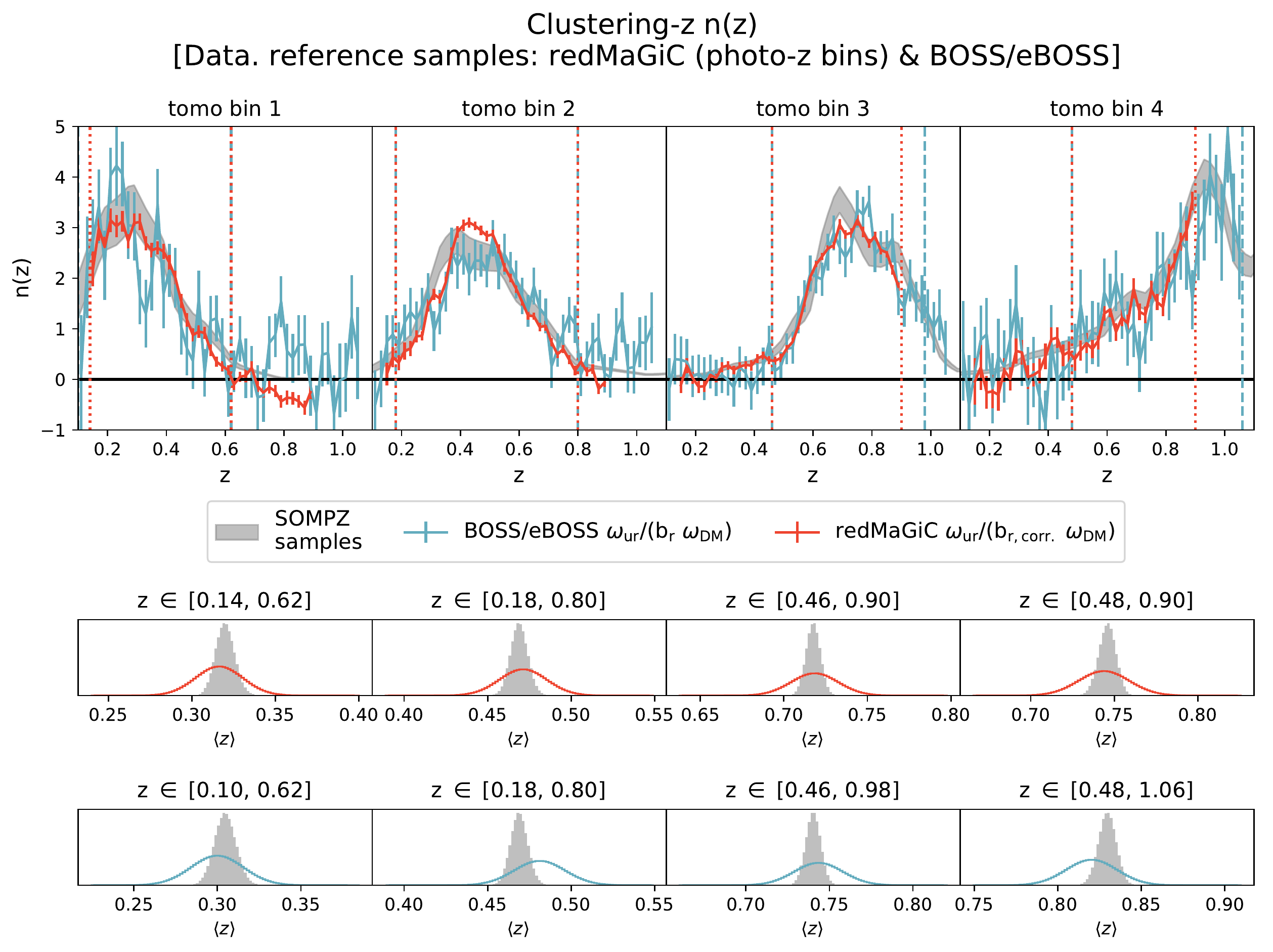}
\end{center}
\caption{Illustration of the agreement among the SOMPZ $n(z)$ and the WZ $n(z)$ obtained using the naive redshift estimator and \redmagic\ and BOSS/eBOSS galaxies as a reference sample. \textit{Upper panels}: the naive redshift distributions estimated per Eq.~\ref{eq:final_estimator} using clustering redshift in data (i.e. no corrections for systematic errors or magnification). \giulia{That is, these are the distributions employed in the mean-matching method.}\gbso{We both show} The redshift distributions obtained using \redmagic\ galaxies as a reference sample, binned using their \redmagic\ photo-$z$ estimates, {are in red. Those}  using BOSS/eBOSS galaxies as a reference sample {are in blue.} The grey bands show the 1-$\sigma$ region encompassed by the SOMPZ realisations.  Vertical dotted (dashed) lines indicate the intervals where the windowed means of the \redmagic\ (BOSS/eBOSS) have been computed.
The lower panels plot the windowed mean redshifts $\avg{z}$ for each bin, as per Eq.~\ref{eq:windowed_mean}, for the two WZ reference samples and for the SOMPZ samples. \marco{The WZ estimates are represented by Gaussian histograms with mean equal to $\langle z \rangle_{\rm WZ}$ and $\sigma$ equal to the uncertainty of the method. The SOMPZ histograms are obtained from the mean redshift of the SOMPZ $n(z)$ realisations.} Good agreement is seen among all three estimators.
}
\label{fig:nz_rmg_eboss_data}
\end{figure*}


{We apply the clustering-$z$ methods to DES Y3 data by first measuring the angle-averaged $w_{\rm ur}(z_i)$ (Eq.~\ref{crosscor2}) of each WL source tomographic bin sample against the \redmagic\ and BOSS/eBOSS samples described in \S\ref{sect:data}.  These cross-correlation data are plotted in Fig.~\ref{fig:w_data}.  Note the exceptionally high $S/N$ level of the \redmagic\ data in particular, even at the rather fine binning of $\Delta z=0.02$ that we use throughout.  Bin-by-bin estimates of the reference bias $b_{\rm r}(z_i)$ are obtained using Eq.~\ref{autocorrelation1}, with a dark-matter $w_{\rm DM}(z_i)$ predicted from theory for nominal cosmological parameters \citep{Planckresults2018}}.

Note that for the \redmagic\ galaxies we
{calculated $b_{\rm r}(z_i)$} applying the correction to the galaxy-matter bias of the reference sample described by Eq. \ref{autocorrref22}, using the fraction of the \redmagic\ galaxies which have a {spectroscopic redshift.}
  \gbso{-z counterpart.} As \redmagic\ galaxies with spec-z counterparts tend to have brighter magnitudes compared to the full \redmagic\ sample, we have applied a magnitude re-weighting to those galaxies before computing the correction, so as to up-weigh (down-weigh) \redmagic\ galaxies under (over) represented in the spec-z subsample. After the re-weighting, the spec-z sample had the same magnitude distribution  of the full \redmagic\ sample. Imperfections in this process should be small based on the tests in previous sections and are included in the systematic uncertainties of the two methods.

\subsection{Mean-matching results}
{We use the mean-matching method as an independent check on the SOMPZ estimates of $n_{\rm u}(z)$ in each tomographic bins.  This begins by calculating the naive (linear-theory, no-magnification, constant-$b_{\rm u}$) redshift distribution $\tilde n_{\rm u}(z_i)$ from Eq.~\ref{eq:final_estimator}, plotted in Fig.~\ref{fig:nz_rmg_eboss_data}.}
\gbso{We present the clustering-based redshift estimates on data in Fig.~\ref{fig:nz_rmg_eboss_data}.}
\gbso{These have been estimated following Eqs.~\ref{system}, \ref{eq:final_estimator}.} We show the distributions obtained with the two reference samples, and, for comparison, the 1-$\sigma$ region encompassed by the SOMPZ realisations.

%

Following the prescription for mean-matching in Eq.~\ref{eq:windowed_mean}, we first compute the mean of the {\redmagic\ and BOSS/eBOSS} clustering-based redshift distributions in the redshift interval where they overlap, also excluding the tails (as detailed at the beginning of Section~\ref{sect:systematic_method1}). We measure differences in $\avg{z}$ \gbso{between the estimates obtained with \redmagic\ galaxies and BOSS/eBOSS galaxies} of $(-0.009 \pm 0.010,\, 0.006 \pm 0.009,\, 0.005 \pm 0.006,\, 0.022 \pm 0.014)$, for the four tomographic bins. The quoted uncertainties take into account the statistical and systematic uncertainties as reported in Table~\ref{table1}, except for the WL galaxy bias uncertainty that is assumed to be shared by the two samples. \marco{The statistical uncertainties are estimated through jackknife resampling. Statistical and systematic uncertainties are added in quadrature.}
{We then compare the $\avg{z}$ values derived for the WZ with two reference samples and the SOMPZ estimates of $n_{\rm u}(z)$: this is shown in the lower panels of Fig.~ \ref{fig:nz_rmg_eboss_data}. 
In this case the full systematic mean-matching uncertainty from Table~\ref{table1} has been included in the WZ values. The WZ values are fully consistent with the SOMPZ values in the mean-matching statistic, although they are weaker.  The behavior is very similar to what was seen in simulations.}




\subsection{Full-shape results}
{Following the procedure used on the simulations, we define a full-shape WZ likelihood using Eqs.~\ref{wursys} and \ref{eq:wzlike}.} We assume fiducial values for the magnification parameters for the \redmagic\ sample, as estimated using Balrog \citep{Suchyta2016,y3-balrog}. We do not have an estimate of the magnification parameters for BOSS/eBOSS galaxies available, so we assumed the same values used for \redmagic\ galaxies. We confirm, however, that assuming values for the magnification parameters that are $-1\times$ the fiducial ones resulted in no relevant effect on the mean of the resultant redshift distributions.  {The nuisance-parameter priors derived from simulations in \S\ref{sect:systematic_method2} are used, including those specifying the allowed variation with $z$ in $b_{\rm u}(z)$ and other elements of the $\sys(z)$ function.}

\begin{figure*}
\begin{center}
\includegraphics[width=1.0 \textwidth]{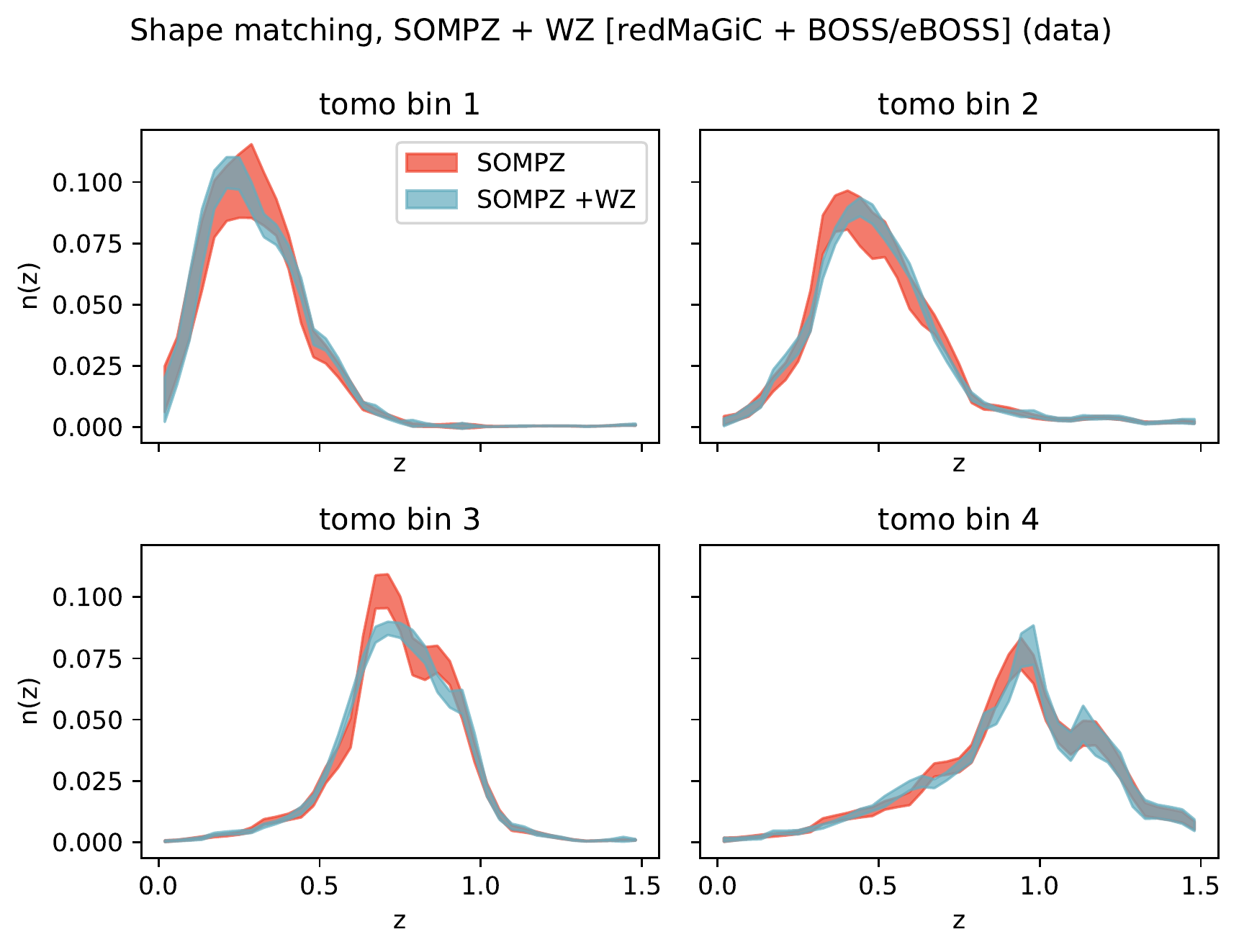}

\end{center}
\caption{SOMPZ redshift distributions, as estimated in data, with and without clustering information (full-shape method). The bands encompass the statistical and systematic uncertainties of the distributions. } 
\label{fig:shape_match_data_all}
\end{figure*}

\marco{Before applying the full-shape method, we checked that the fiducial $\hat{w}_{\rm ur}$ model on data (obtained using SOMPZ $n_{\rm u}(z)$ as baseline) was compatible with the measured $w_{\rm ur}$ marginalised over the systematic function ${\rm Sys(z)}$. This check has been performed separately for \redmagic\ and BOSS/eBOSS.} {We then use the Hamiltonian Monte Carlo method to draw samples from the joint posterior distribution of the SOMPZ likelihood and the WZ likelihoods for both \redmagic\ and BOSS/eBOSS WZ data. Fig.~\ref{fig:shape_match_data_all} show the 68\% confidence interval of the $n_{\rm u}(z)$ samples from the SOMPZ$+$WZ posterior, as well as those from the pure SOMPZ posterior. At redshifts where WZ information is available, it greatly reduces the point-by-point uncertainties in $n_{\rm u}(z)$, just as in the simulations.  The WZ full-shape method is thus very successful at reducing the impact sample variance on SOMPZ estimators.}

{The averages and standard deviations of the mean-$z$ distributions of the SOMPZ and SOMPZ$+$WZ posteriors are listed in Table~\ref{tableres_data}, along with the results of importance-weighting the SOMPZ samples with the mean-matching likelihood in Eq.~\ref{likeY1}.  As expected from the simulations, the WZ information does not substantially alter the bin means derived from photo-$z$ methods, in both the mean-matching and full-shape methods. The significant improvement in shape accuracy, as seen in Fig.~\ref{fig:shape_match_data_all}, is the principal product of the correlation-redshift method for DES Y3 analyses.
}

 
\begin{table*}

\caption {\textbf{Data}. The mean redshift estimates of the SOMPZ distributions with and without clustering-based information.}
\centering
\begin{adjustbox}{width=0.9\textwidth}
\begin{tabular}{|c|c|c|c|c|}
\hline

\textbf{case} & \textbf{tomo bin 1}& \textbf{tomo bin 2} & \textbf{tomo bin 3}& \textbf{tomo bin 4}\\
\hline

SOMPZ $\langle z \rangle$:	& $0.318 \pm 0.009$ & $0.513 \pm 0.006$ & $0.750 \pm 0.005$ & $ 0.942 \pm 0.011$ \\
SOMPZ $+$ WZ (mean-matching) :	& $0.317 \pm 0.008$ & $ 0.514 \pm 0.006$ & $ 0.750 \pm 0.005$ & $ 0.941 \pm 0.011$ \\
SOMPZ $+$ WZ (full-shape) : & $0.321 \pm 0.008$ & $0.517 \pm 0.006$ & $0.749 \pm 0.005$ & $ 0.940 \pm 0.010$ \\

\hline
\end{tabular}
\end{adjustbox}
\label{tableres_data}
\end{table*}

\section{Conclusions}\label{sect:conclusions}

This work describes the use of clustering measurements to constrain the weak-lensing (WL) source galaxy redshift distributions for the Dark Energy Survey Year 3 (DES Y3) cosmological analyses.
We cross-correlate the WL source galaxies (the ``unknown'' sample $u$) with ``reference'' samples ($r$) from both the DES Y3 \redmagic\ catalog (luminous red galaxies with secure photometric redshifts) and BOSS/eBOSS galaxies (with spec-$z$ estimates). The reference samples are divided into thin redshift bins centered at $\{z_i\}$ to yield 2-point angular cross-correlation measurements $w_{\rm ur}(z_i),$ for each combination of reference sample and WL tomographic bin, following now-standard practices for clustering-redshift (WZ) methods.  The $w_{\rm ur}(z_i)$ measurements are weighted over angular separation to maximize the overall $S/N$ ratio while avoiding the large angular scales used for cosmological measurements, in order to keep the WZ inferences statistically independent of the cosmological data vectors.

We describe two distinct methods to constrain the redshift distributions $n_{\rm u}(z)$ of the unknown samples using the $w_{\rm ur}(z_i)$ data.  The ``mean-matching'' method focuses on the mean $\avg{z}$ of the redshift distribution over
a redshift window bounded by the redshift range of the reference sample and the 2-$\sigma$ extent of $n_{\rm u}(z)$.  This method, similar to what was used in DES Y1 analyses (\citealt*{Gatti2018}, \citealt{Davis2018}), starts by computing the $\avg{z}$ of a naive WZ estimate $\tilde n_{\rm u}(z_i)$ (per Eq.~\ref{eq:final_estimator}) that assumes linear biasing with constant $b_{\rm u}$ and no magnification.  From simulations, we estimate the additional uncertainty on $\avg{z}$ that arises from systematic errors in the naive estimator, which we conservatively take as $0.014$ and are dominated by the unknown redshift dependence of $b_{\rm u}(z).$  Finally we can compare this WZ estimate of $\avg{z}$ to that of the $n_{\rm u}(z)$ inferred from photo-$z$ or some other independent method.  For the DES Y3 data, we find the mean-matching method indicates full consistency between the SOMPZ photometric estimator and the WZ estimators, for all combinations of tomographic bin and reference sample.

The systematic errors we derive on $\avg{z}$ are larger than the statistical errors (estimated through jackknife) with DES Y3 data (Table~\ref{table1}), and also larger than the total errors estimated for the SOMPZ method (Fig.~\ref{fig:nz_rmg_eboss_data}).  Thus this mean-matching approach has reached the limits of its usefulness, unless future experiments obtain narrower WL tomographic bins, and/or obtain external information on the relative bias of the unknown sample against the reference samples. Indeed the degeneracy between $n_{\rm u}(z)$ and $b_{\rm u}(z)$ in the observable $w_{\rm ur}(z)$ is the fundamental limitation of the WZ approach.
This does not, however, mean that we have exhausted the information available from the WZ data in general.  As discussed at the end of \S\ref{sect:systematic_method1}, the mean $z$ is probably the summary statistic of $n_{\rm u}(z)$ that is \emph{most} degraded by the dominant systematic error, redshift-dependent bias $b_{\rm u}(z),$  because this unmodelled multiplicative contribution to $w_{\rm ur}(z)$ is a smooth function of redshift.  Higher-order moments, or more generally the detailed shape of $n_{\rm u}(z),$ are less susceptible to WZ systematic errors, which are all expected to be smooth functions of $z$.

To extract this information, we apply the ``full-shape'' method, developed in section \ref{sect:systematic_method2}, using $w_{\rm ur}(z)$ data to inform $n_{\rm u}(z)$.  We allow our model $\hat w_{\rm ur}(z)$ to incorporate an arbitrary multiplicative function $\sys(z,\vecs)$ of redshift and nuisance parameters $\vecs$ that will mimic the effects of smooth systematic errors, such as $b_{\rm u}(z)$ and failure of the linear-bias model at small scales.  Using simulations to choose this function and an appropriately flexible prior on $\vecs,$ we can now define a WZ likelihood for an arbitrary choice of $n_{\rm u}(z)$ that marginalizes over these nuisance parameters, as well as nuisances associated with lensing magnification signals that contaminate $w_{\rm ur}(z)$ (Eq.~\ref{eq:wzlike}).


We note that this method improves on previous applications of WZ methods to WL cosmology (e.g., \citealt{Choi2016,Davis2017aaa,vandenBusch2020}), which mostly have used the former to constrain simple shifts of photo-$z$-derived redshift distributions, i.e. $n_{\rm u}(z)=n_{\rm PZ}(z + \Delta z)$. These approaches can lead to biased results if the shape of the photo-$z$ posterior differs from the truth, or if clustering  systematics are not taken into account by a sufficiently flexible model, as noted by \citealt*{Gatti2018}; \citealt{vandenBusch2020}. We improve on these approaches for DES Y3 by defining likelihoods for $n_{\rm u}(z)$ arising from both SOMPZ and WZ methods, and using Hamiltonian Monte Carlo (HMC) to sample $n_{\rm u}(z)$ realizations from the product of these independent likelihoods.  This also allows us to combine the information of the \redmagic\ and BOSS/eBOSS references into a single inference.  Note that each sample of the chain specifies redshift distributions for all four tomographic bins, capturing any inter-bin correlations that arise from the SOMPZ inference.  This SOMPZ$+$WZ technique is extremely successful at reducing the point-by-point uncertainty in $n_{\rm PZ}(z)$ that arises from sample variance in the small surveys typically used to calibrate photo-$z$ methods.  The results for DES Y3 data can be seen in Figure~\ref{fig:shape_match_data_all}.  The addition of the full-shape WZ information to SOMPZ yields $n_{\rm u}(z)$ samples that are much smoother and more realistic, taking advantage of the very high $S/N$ that we have in $w_{\rm ur}(z)$ from the full footprint of DES Y3.  This benefit is present despite the fact that the full-shape method does little to improve the SOMPZ's estimate of the mean redshift of each bin.

\marco{The final DES Y3 redshift calibration strategy includes a few additional minor tweaks to the SOMPZ$+$WZ samples, not addressed here.  The $n(z)$ realisations are modified to account for uncertainties in the photometric calibration of the SOMPZ inputs, and the $z<0.1$ behavior (which is not constrained by WZ data) is smoothed to a physically reasonable form.  {These steps mostly affect the first tomographic bin \citep*{y3-sompz}} An additional correction to all the $n(z)$ realisations is performed to account for the effects of blending, based on the work on image simulations described in \cite{MacCrannSims2019}. Then, ideally, the realisations are sampled over during the cosmological analysis, using the \textit{hyperrank} technique \citep{y3-hyperrank}. In practice, however, in our fiducial cosmological run, we decided to parametrise the $n(z)$ uncertainties by shifts around their mean with a shift parameter $\Delta z$. This choice was dictated by efficiency reasons, and by the fact that we verified in \citet{y3-hyperrank} that marginalising over the mean of the redshift distributions rather than sampling over the multiple $n(z)$ realisations was sufficient for the DES Y3 analysis. The prior on $\Delta z$ is naturally provided by the scatter on the mean of the $n(z)$ realisations. Finally, when sampling the cosmological parameters, further constraints on the $n(z)$ are provided by the ``shear-ratio'' test \citep{y3-shearratio}. The shear-ratio test uses small-scale galaxy-galaxy lensing measurements to further inform the shifts $\Delta z$. In practice, when running the cosmological analysis, the shear-ratio likelihood is simply multiplied by the cosmological likelihood, since the two are independent. Having combined these sources of information on $n(z),$ we find in \cite{y3-3x2ptkp} that its uncertainties are insignificant contributors to the Y3 cosmological uncertainty, despite these data having the smallest statistical uncertainties of any photometric cosmology survey to date.}




The techniques used in this paper are applicable to other large imaging surveys.  Further improvements in accuracy could be possible from having a reference sample that has spectroscopic redshifts like BOSS/eBOSS (eliminating one systematic error source) but large area and very high $S/N$ like the DES Y3 \redmagic\ sample.  Improved prior knowledge of the magnification coefficients $\alpha_{\rm u},\alpha_{\rm r}$ would also be of use.  Importantly, the impact of bias evolution on WZ measures scales as $(\Delta z)^2,$ where $\Delta z$ is the rough width of each tomographic source bin, so improved binning accuracy from photo-$z$'s will increase the value of clustering redshifts. Ultimately the scheme of (\citealt{Sanchez2019}; \citealt*{Alarcon2019}) whereby one samples the posterior of the actual mass density field, individual source $z's$, and bias functions as constrained by the full catalogs, may offer stronger information than WZ methods that reduce the catalogs to the summary 2-point statistics $w_{\rm ur}(z).$  But the methods applied to DES Y3 do make more complete use of the WZ data at summary-statistic level than has been done in the past.

\section*{Data Availability}
The full \mcal\ and \redmagic\ catalogues will be made publicly available following publication, at the URL \url{https://des.ncsa.illinois.edu/releases}. The code used to perform the tests in this manuscript will be made available upon reasonable request to the authors.

\section*{Acknowledgements}

The project leading to these results have received funding from ''la Caixa'' Foundation (ID 100010434), under the fellowship LCF/BQ/DI17/11620053 and has received funding from the European Union's Horizon 2020 research and innovation programme under the Marie Sk\l{}odowska-Curie grant agreement No. 713673.

Funding for the DES Projects has been provided by the U.S. Department of Energy, the U.S. National Science Foundation, the Ministry of Science and Education of Spain, 
the Science and Technology Facilities Council of the United Kingdom, the Higher Education Funding Council for England, the National Center for Supercomputing 
Applications at the University of Illinois at Urbana-Champaign, the Kavli Institute of Cosmological Physics at the University of Chicago, 
the Center for Cosmology and Astro-Particle Physics at the Ohio State University,
the Mitchell Institute for Fundamental Physics and Astronomy at Texas A\&M University, Financiadora de Estudos e Projetos, 
Funda{\c c}{\~a}o Carlos Chagas Filho de Amparo {\`a} Pesquisa do Estado do Rio de Janeiro, Conselho Nacional de Desenvolvimento Cient{\'i}fico e Tecnol{\'o}gico and 
the Minist{\'e}rio da Ci{\^e}ncia, Tecnologia e Inova{\c c}{\~a}o, the Deutsche Forschungsgemeinschaft and the Collaborating Institutions in the Dark Energy Survey.

The Collaborating Institutions are Argonne National Laboratory, the University of California at Santa Cruz, the University of Cambridge, Centro de Investigaciones Energ{\'e}ticas, 
Medioambientales y Tecnol{\'o}gicas-Madrid, the University of Chicago, University College London, the DES-Brazil Consortium, the University of Edinburgh, 
the Eidgen{\"o}ssische Technische Hochschule (ETH) Z{\"u}rich, 
Fermi National Accelerator Laboratory, the University of Illinois at Urbana-Champaign, the Institut de Ci{\`e}ncies de l'Espai (IEEC/CSIC), 
the Institut de F{\'i}sica d'Altes Energies, Lawrence Berkeley National Laboratory, the Ludwig-Maximilians Universit{\"a}t M{\"u}nchen and the associated Excellence Cluster Universe, 
the University of Michigan, the National Optical Astronomy Observatory, the University of Nottingham, The Ohio State University, the University of Pennsylvania, the University of Portsmouth, 
SLAC National Accelerator Laboratory, Stanford University, the University of Sussex, Texas A\&M University, and the OzDES Membership Consortium.

Based in part on observations at Cerro Tololo Inter-American Observatory at NSF's NOIRLab (NOIRLab Prop. ID 2012B-0001; PI: J. Frieman), which is managed by the Association of Universities for Research in Astronomy (AURA) under a cooperative agreement with the National Science Foundation.

The DES data management system is supported by the National Science Foundation under Grant Numbers AST-1138766 and AST-1536171.
The DES participants from Spanish institutions are partially supported by MINECO under grants AYA2015-71825, ESP2015-66861, FPA2015-68048, SEV-2016-0588, SEV-2016-0597, and MDM-2015-0509, 
some of which include ERDF funds from the European Union. IFAE is partially funded by the CERCA program of the Generalitat de Catalunya.
Research leading to these results has received funding from the European Research
Council under the European Union's Seventh Framework Program (FP7/2007-2013) including ERC grant agreements 240672, 291329, and 306478.
We  acknowledge support from the Brazilian Instituto Nacional de Ci\^encia
e Tecnologia (INCT) e-Universe (CNPq grant 465376/2014-2).

This manuscript has been authored by Fermi Research Alliance, LLC under Contract No. DE-AC02-07CH11359 with the U.S. Department of Energy, Office of Science, Office of High Energy Physics.

Funding for the Sloan Digital Sky Survey IV has been provided by the Alfred P. Sloan Foundation, the U.S. Department of Energy Office of Science, and the Participating Institutions. SDSS acknowledges support and resources from the Center for High-Performance Computing at the University of Utah. The SDSS web site is www.sdss.org.

SDSS is managed by the Astrophysical Research Consortium for the Participating Institutions of the SDSS Collaboration including the Brazilian Participation Group, the Carnegie Institution for Science, Carnegie Mellon University, Center for Astrophysics | Harvard \& Smithsonian (CfA), the Chilean Participation Group, the French Participation Group, Instituto de Astrofisica de Canarias, The Johns Hopkins University, Kavli Institute for the Physics and Mathematics of the Universe (IPMU) / University of Tokyo, the Korean Participation Group, Lawrence Berkeley National Laboratory, Leibniz Institut f{\"u}r Astrophysik Potsdam (AIP), Max-Planck-Institut f{\"u}r Astronomie (MPIA Heidelberg), Max-Planck-Institut f{\"u}r Astrophysik (MPA Garching), Max-Planck-Institut f{\"u}r Extraterrestrische Physik (MPE), National Astronomical Observatories of China, New Mexico State University, New York University, University of Notre Dame, Observatorio Nacional / MCTI, The Ohio State University, Pennsylvania State University, Shanghai Astronomical Observatory, United Kingdom Participation Group, Universidad Nacional Aut\'{o}noma de M\'{e}xico, University of Arizona, University of Colorado Boulder, University of Oxford, University of Portsmouth, University of Utah, University of Virginia, University of Washington, University of Wisconsin, Vanderbilt University, and Yale University.



\appendix
\section{Full $\hat{w}_{\rm ur}$ model and analytical marginalisation}\label{AppendixSys}

We provide here more details about the implementation of the full-shape method. The method assigns a likelihood (Eq.~\ref{eq:wzlike}) of the observed $w_{\rm ur}(z_i)$ given a proposal for the
redshift distributions $\{n_{\rm u}(z_i)\}$ along with a set of other relevant parameters. The likelihood uses the model in Eq.~\ref{wursys}.  We will assume that the values of the dark-matter correlation $w_{\rm DM}(z_i),$ the reference-sample properties $b_{\rm r}(z)$ and $\alpha_{\rm r}(z)$, and the magnification coefficients $D_{ij}$ are provided along with $n_{\rm u}(z_i)$.  We will consider as nuisance parameters the properties of the unknown population, namely the $\alpha_{\rm u}(z)$ and $b_{\rm u}(z)$ used in magnification terms; plus any parameters \vecs\ of the $\sys(z)$ function that allows for systematic errors.

We will assume here that $\alpha_{\rm u}$ and $b_{\rm u}$ are independent of redshift, though in principle a more general function, linear in some parameters, can be used without altering any of the methods herein. We note that we did not multiply the magnification terms by the systematic function: despite the fact that the magnification terms are not immune to systematic errors, we assumed that it was not necessary to further modelling those, as the $\alpha_{\rm r}$, $\alpha_{\rm u}$ and $\bar{b}_{\rm u}$ parameters provide enough flexibility to the model and the magnification signal is much smaller than clustering to start with.  We also note that the  $b_{\rm u}$ parameter is used only in the magnification term, and hence can be independent of the bias for clustering that is absorbed into the $\sys(z)$ function.  This allows for the systematic errors in the magnification term to differ from those in the clustering term.

The systematic-error function for clustering is given the exponentiated polynomial form in Eqs.~\ref{eq:model_syst_unc} and \ref{eq:sys_rescale}. Tuning the order $M$ allows us to adjust the smoothness of the function, and exponentiation allows us to draw the coefficients $\vecs$ from 1-d Gaussian priors while maintaining positive $\sys(z)$.  Adjusting the $\sigma_s$ values of these priors tunes the RMS of the systematic variations, in a way made predictable by the orthogonality of the Legendre polynomials.  \gary{We wish for independent, uniform Gaussian priors on the $s_i$ to propagate into RMS variation of $\log\sys(z)$ that is approximately independent of $z$ over $[z_{\rm min},z_{\rm max}].$   The Legendre polynomials have this property over most of their nominal domain $u\in [-1,1],$ but not near the edges of this range.  For this reason we map $[z_{\rm min},z_{\rm max}]\rightarrow[-0.85,0.85],$ as indicated by Eq.~\ref{eq:sys_rescale}.}

Eq.~\ref{eq:wzlike} requires us to marginalize over the nuisance-parameter vector $\vecq=\{\vecp\,\vecs\}$ (with $\vecp = {b_{\rm u},\alpha_{\rm u}u}$). Doing so as part of a Markov chain would be unweildly, introducing 8 free parameters for each of the 4 tomographic bins times 2 reference samples.  It is far better to execute the marginalization on the fly during sampling if possible.  The log-likelihood is not quite quadratic in \vecq---the exponentiation of the polynomial in $\sys(z;\vecs)$ makes the model $\hat w_{\rm ur}$ non-linear in \vecs.  We opt to linearize the model about its \marco{maximum $\vecs_0 = \{s_{k,0}\}$}: 
\begin{multline}
\sys\left(z_i,\vecs\right) \approx \sys\left(z_i,\vecs_0\right) \times \\ \left[ 1 + \sum_{k=0}^{M} \frac{\sqrt{2k+1}}{0.85} P_k(u)s_{k,0}\left(s_k-s_{k,0}\right)\right], 
\label{eq:linearize}
\end{multline}
The deviation of the data from the model can then be rewritten in linear form, with ${\bf w}_{\rm ur}$ being a vector over redshifts, as 
\begin{equation}
{\bf w}_{\rm ur}-\hat{\bf w}_{\rm ur} = \mathbf{c}(\vecq_0) - A \vecq
\end{equation}
where $\mathbf{c}$ is a vector independent of \vecq\ and $A$ is a matrix composed of the linear terms in Eq.~\ref{eq:linearize} and elements of the magnification terms.

If we assume the nuisance parameters we want to marginalise over to have a Gaussian prior $\vecq \sim \mathcal{N}(\mathbf{\mu}_q,\Sigma_q)$ , we can write the full likelihood as follows:
\begin{multline}
\mathcal{L_{\rm WZ}} \approx \vert 2 \pi \Sigma_{\mathrm{wz}} \vert^{-1/2} \vert 2 \pi \Sigma_{p} \vert^{-1/2}\, \times\\ \int  d \vecq \, \exp\left[- \frac{1}{2} (\mathbf{c} - A \vecq)^T \hat{\Sigma}_{\mathrm{wz}}^{-1}(\mathbf{c}-A\vecq)\right] \times \\ \exp \left[- \frac{1}{2} (\vecq-\mathbf{\mu}_q)^T \hat{\Sigma}_q^{-1}(\vecq-\mathbf{\mu}_q) \right],
\end{multline}
This is a Gaussian integral that can be reduced to linear algebra.

In summary, the algorithm for the marginalization in Eq.~\ref{eq:wzlike} is:
\begin{enumerate}
    \item Find the values $\vecq_0$ which maximize the integrand.  This is done using Newton iterations.
    \item Evaluate the vector $\mathbf{c}$ and matrix $A$ at this value of $\vecq_0.$
    \item Substitute these and the known $\Sigma_{\rm wz}, \mathbf{\mu}_q,$ and $\Sigma_q$ into the analytic result for the Gaussian integral above.
\end{enumerate}
Although this marginalization is approximate, it does not actually need to be exact, \gary{because the chosen functional form for $\sys(z,\vecs)$ is somewhat arbitrary.}  All that is necessary is that the algorithm yields a \gary{likelihood $\mathcal{L}$ of the WZ data given a proposed $n_{\rm u}(z)$ that decreases in a meaningful way and robust way as the data move away from the naive linear model.}



\bibliographystyle{mn2e_2author_arxiv_amp.bst}
\bibliography{references}

\bsp	
\label{lastpage}

\section*{Affiliations}
$^{1}$ Institut de F\'{\i}sica d'Altes Energies (IFAE), The Barcelona Institute of Science and Technology, Campus UAB, 08193 Bellaterra (Barcelona) Spain\\	
$^{2}$ Department of Physics and Astronomy, University of Pennsylvania, Philadelphia, PA 19104, USA\\
$^{3}$ Argonne National Laboratory, 9700 South Cass Avenue, Lemont, IL 60439, USA \\
$^{4}$ Department of Physics, Stanford University, 382 Via Pueblo Mall, Stanford, CA 94305, USA \\
$^{5}$ Kavli Institute for Particle Astrophysics \& Cosmology, P. O. Box 2450, Stanford University, Stanford, CA 94305, USA \\
$^{6}$ Physics Department, 2320 Chamberlin Hall, University of Wisconsin-Madison, 1150 University Avenue Madison, WI  53706-1390 \\
$^{7}$ Department of Physics, Duke University Durham, NC 27708, USA \\
$^{8}$ Department of Astronomy, University of California, Berkeley,  501 Campbell Hall, Berkeley, CA 94720, USA \\
$^{9}$ Santa Cruz Institute for Particle Physics, Santa Cruz, CA 95064, USA\\
$^{10}$ Center for Cosmology and Astro-Particle Physics, The Ohio State University, Columbus, OH 43210, USA \\
$^{11}$ SLAC National Accelerator Laboratory, Menlo Park, CA 94025, USA \\
$^{12}$ Department of Physics, The Ohio State University, Columbus, OH 43210, USA\\
$^{13}$ Jodrell Bank Center for Astrophysics, School of Physics and Astronomy, University of Manchester, Oxford Road, Manchester, M13 9PL, UK\\
$^{14}$ Department of Astronomy and Astrophysics, University of Chicago, Chicago, IL 60637, USA\\
$^{15}$ Fermi National Accelerator Laboratory, P. O. Box 500, Batavia, IL 60510, USA\\
$^{16}$ Institut d'Estudis Espacials de Catalunya (IEEC), 08034 Barcelona, Spain\\
$^{17}$ Institute of Space Sciences (ICE, CSIC),  Campus UAB, Carrer de Can Magrans, s/n,  08193 Barcelona, Spain\\
$^{18}$ Department of Physics, University of Arizona, Tucson, AZ 85721, USA\\
$^{19}$ Cerro Tololo Inter-American Observatory, NSF's National Optical-Infrared Astronomy Research Laboratory, Casilla 603, La Serena, Chile\\
$^{20}$ Departamento de F\'isica Matem\'atica, Instituto de F\'isica, Universidade de S\~ao Paulo, CP 66318, S\~ao Paulo, SP, 05314-970, Brazil\\
$^{21}$ Laborat\'orio Interinstitucional de e-Astronomia - LIneA, Rua Gal. Jos\'e Cristino 77, Rio de Janeiro, RJ - 20921-400, Brazil\\
$^{22}$ Instituto de Fisica Teorica UAM/CSIC, Universidad Autonoma de Madrid, 28049 Madrid, Spain\\
$^{23}$ Institute of Cosmology and Gravitation, University of Portsmouth, Portsmouth, PO1 3FX, UK\\
$^{24}$ CNRS, UMR 7095, Institut d'Astrophysique de Paris, F-75014, Paris, France\\
$^{25}$ Sorbonne Universit\'es, UPMC Univ Paris 06, UMR 7095, Institut d'Astrophysique de Paris, F-75014, Paris, France\\
$^{26}$ Department of Physics \& Astronomy, University College London, Gower Street, London, WC1E 6BT, UK\\
$^{27}$ Instituto de Astrofisica de Canarias, E-38205 La Laguna, Tenerife, Spain\\
$^{28}$ Universidad de La Laguna, Dpto. Astrofísica, E-38206 La Laguna, Tenerife, Spain\\
$^{29}$ Department of Astronomy, University of Illinois at Urbana-Champaign, 1002 W. Green Street, Urbana, IL 61801, USA\\
$^{30}$ National Center for Supercomputing Applications, 1205 West Clark St., Urbana, IL 61801, USA\\
$^{31}$ University of Nottingham, School of Physics and Astronomy, Nottingham NG7 2RD, UK\\

$^{32}$ INAF-Osservatorio Astronomico di Trieste, via G. B. Tiepolo 11, I-34143 Trieste, Italy\\
$^{33}$ Institute for Fundamental Physics of the Universe, Via Beirut 2, 34014 Trieste, Italy\\
$^{34}$ Observat\'orio Nacional, Rua Gal. Jos\'e Cristino 77, Rio de Janeiro, RJ - 20921-400, Brazil\\
$^{35}$ Department of Physics, University of Michigan, Ann Arbor, MI 48109, USA\\
$^{36}$ Department of Physics and Astronomy, University of Utah, 115 S. 1400 E., Salt Lake City, UT 84112, USA\\
$^{37}$ Department of Physics, IIT Hyderabad, Kandi, Telangana 502285, India\\
$^{38}$ Department of Astronomy/Steward Observatory, University of Arizona, 933 North Cherry Avenue, Tucson, AZ 85721-0065, USA\\
$^{39}$ Jet Propulsion Laboratory, California Institute of Technology, 4800 Oak Grove Dr., Pasadena, CA 91109, USA\\
$^{40}$ Department of Astronomy, University of Michigan, Ann Arbor, MI 48109, USA\\
$^{41}$ Institute of Theoretical Astrophysics, University of Oslo. P.O. Box 1029 Blindern, NO-0315 Oslo, Norway\\
$^{42}$ IKavli Institute for Cosmological Physics, University of Chicago, Chicago, IL 60637, USA\\
$^{43}$ Institute of Astronomy, University of Cambridge, Madingley Road, Cambridge CB3 0HA, UK\\
$^{44}$ Kavli Institute for Cosmology, University of Cambridge, Madingley Road, Cambridge CB3 0HA, UK\\
$^{45}$ School of Mathematics and Physics, University of Queensland,  Brisbane, QLD 4072, Australia\\
$^{46}$ Faculty of Physics, Ludwig-Maximilians-Universit\"at, Scheinerstr. 1, 81679 Munich, Germany\\
$^{47}$ Max Planck Institute for Extraterrestrial Physics, Giessenbachstrasse, 85748 Garching, Germany\\
$^{48}$ Universit\"ats-Sternwarte, Fakult\"at f\"ur Physik, Ludwig-Maximilians Universit\"at M\"unchen, Scheinerstr. 1, 81679 M\"unchen, Germany\\
$^{49}$ Center for Astrophysics $\vert$ Harvard \& Smithsonian, 60 Garden Street, Cambridge, MA 02138, USA\\
$^{50}$ Australian Astronomical Optics, Macquarie University, North Ryde, NSW 2113, Australia\\
$^{51}$ Lowell Observatory, 1400 Mars Hill Rd, Flagstaff, AZ 86001, USA\\
$^{52}$ Department of Applied Mathematics and Theoretical Physics, University of Cambridge, Cambridge CB3 0WA, UK\\
$^{53}$ George P. and Cynthia Woods Mitchell Institute for Fundamental Physics and Astronomy, and Department of Physics and Astronomy, Texas A\&M University, College Station, TX 77843,  USA\\
$^{54}$ Department of Astrophysical Sciences, Princeton University, Peyton Hall, Princeton, NJ 08544, USA\\
$^{55}$ Instituci\'o Catalana de Recerca i Estudis Avan\c{c}ats, E-08010 Barcelona, Spain\\
$^{56}$ Waterloo Centre for Astrophysics, University of Waterloo, 200 University Ave W, Waterloo, ON N2L 3G1, Canada\\
$^{57}$ Department of Physics and Astronomy, University of Waterloo, 200 University Ave W, Waterloo, ON N2L 3G1, Canada\\
$^{58}$ Perimeter Institute for Theoretical Physics, 31 Caroline St. North, Waterloo, ON N2L 2Y5, Canada\\
$^{59}$ Centro de Investigaciones Energ\'eticas, Medioambientales y Tecnol\'ogicas (CIEMAT), Madrid, Spain\\
$^{60}$ Department of Physics and Astronomy, Sejong University, Seoul, 143-747, Korea\\
$^{61}$ Department of Physics, Carnegie Mellon University, Pittsburgh, Pennsylvania 15312, USA\\
$^{62}$ School of Physics and Astronomy, University of Southampton,  Southampton, SO17 1BJ, UK \\
$^{63}$ Computer Science and Mathematics Division, Oak Ridge National Laboratory, Oak Ridge, TN 37831\\
$^{64}$ Department of Physics and Astronomy, Pevensey Building, University of Sussex, Brighton, BN1 9QH, UK\\

\end{document}